\documentclass[useAMS,usenatbib]{mn2e}
\usepackage{amsmath,psfig,epsfig,graphics}

\def\aap{A\&A}

\def\apj{ApJ}

\def\apjl{ApJ}
\def\mnras{MNRAS}
\def\araa{ARA\&A}

\def\apjs{ApJS}

\def\ssr{Space Sci. Rev. }
\def\pasj{PASJ}

\def\lesssim{\mathrel{\hbox{\rlap{\hbox{\lower4pt\hbox{$\sim$}}}\hbox{$<$}}}}
\def\gesssim{\mathrel{\hbox{\rlap{\hbox{\lower4pt\hbox{$\sim$}}}\hbox{$>$}}}}

\def\lesssim{\mathrel{\hbox{\rlap{\hbox{\lower4pt\hbox{$\sim$}}}\hbox{$<$}}}}
\def\gesssim{\mathrel{\hbox{\rlap{\hbox{\lower4pt\hbox{$\sim$}}}\hbox{$>$}}}}

\begin{document}

\author[Morandi et al.]
{Andrea Morandi${}^1$\thanks{E-mail: andrea.morandi@uah.edu}, Ming Sun${}^1$, William Forman${}^2$, Christine Jones${}^2$\\
$^{1}$ Physics Department, University of Alabama in Huntsville, Huntsville, AL 35899, USA\\
$^{2}$ Harvard-Smithsonian Center for Astrophysics, 60 Garden St., Cambridge, MA 02138, USA\\
}

\title[The galaxy cluster outskirts probed by {\em Chandra}]
{The galaxy cluster outskirts probed by {\em Chandra}}
\maketitle

\begin{abstract}
We studied the physical properties of the intracluster medium in the virialization region of a sample of 320 clusters ($0.056 <z< 1.24$, $kT\gesssim 3$ keV) in the {\em Chandra} archive. With the emission measure profiles from this large sample, the typical gas density, gas slope and gas fraction can be constrained out to and beyond $R_{200}$. We observe a steepening of the density profiles beyond $R_{500}$ with $\beta \sim 0.68$ at $R_{500}$ and $\beta \sim 1$ at $R_{200}$ and beyond. By tracking the direction of the cosmic filaments approximately with the ICM eccentricity, we report that galaxy clusters deviate from spherical symmetry, with only small differences between relaxed and disturbed systems. We also did not find evolution of the gas density with redshift, confirming its self-similar evolution. The value of the baryon fraction reaches the cosmic value at $R_{200}$: however, systematics due to non-thermal pressure support and clumpiness might enhance the measured gas fraction, leading to an actual deficit of the baryon budget with respect to the primordial value. This study has important implications for understanding the ICM physics in the outskirts.
\end{abstract}


\section{Introduction}\label{intro}
Exploring the virialization region of galaxy clusters is vital to our understanding of large scale structure formation, offering a direct view of cluster growth. This topic has recently raised the attention of the scientific community, since an accurate measurement of the mass and baryonic fraction of cluster outskirts can provide constraints on cosmological parameters, and allow us to test the validity of the CDM framework \citep[e.g.][]{reiprich2013}.

Nevertheless, the outskirts remain a relatively unexplored territory, bearing the signature of a complex physics such as ongoing accretion processes, significant departures from virialization and hydrostatic equilibrium \citep{lau2009}, clumping of the gas \citep[][]{morandi2014}. These effects may bias, in turn, the measurements of cluster masses, thus limiting the use of galaxy clusters as a high-precision cosmological proxy. Hence a deeper understanding of the state of the intracluster gas in cluster outskirts is required. 

Observations are very challenging in cluster outskirts, since the X-ray surface brightness drops below the background level at large radii. Thanks to its low particle background from its low orbit, {\em Suzaku} observations have extended X-ray measurements of the intracluster medium (ICM) profiles out to and beyond $R_{200}$, where $R_{200}$ is the radius within which the mean total density is 200 times the critical density of the Universe. Initial results were surprising, revealing significant departures from the theoretical predictions. These observations showed that, at large cluster-centric radii ($R\gesssim R_{200}$), the observed entropy profiles becomes flatter \citep[e.g.,][]{bautz2009,kawaharada2010}, and the observed gas fraction exceeded the cosmic baryon fraction of the universe \citep{simionescu2011}. {\em Suzaku} measurements call for appreciable gas clumpiness in the outskirts, inferred from its effects on thermodynamic profiles \citep{walker2013} or from the high values, larger than the cosmic baryon fraction, of the gas mass fraction towards the virial radius. However, difficulties arise in modeling out point sources and/or galactic foreground with the limited angular resolution ($\sim 2$ arcmin) of {\em Suzaku} at the level required for measuring the extremely low surface brightness in the cluster outskirts. In particular, the {\em Suzaku} point-spread function and the stray light can cause significant contamination from the brightness in the central regions to large radii. Moreover, {\em Suzaku} observations have been mostly performed along narrows arms, which might reflects preferential directions connected with the large-scale structure (e.g., in the direction of filaments). Therefore, it is possible that the aforementioned measurements are not representative of the cluster as a whole. 

Independent {\em ROSAT} PSPC observations (with low background and large field of view) have shown the steepening of the gas density profile at large radii, but at the level considerably larger than those inferred from the {\em Suzaku} observations \citep[the ICM slope $\beta=0.890 \pm 0.026$ at $R_{200}$,][]{eckert2012}. 

In this respect, independent measurements with {\em Chandra} with respect to {\em Suzaku} can shed light on systematic uncertainties in the current measurements \citep[e.g.,][]{ettori2009a,ettori2011,moretti2011}: indeed, despite its higher particle background, {\em Chandra} provides a superior angular resolution which is fundamental in order to: i) remove emission from unrelated sources, e.g. point sources; ii) to constrain the emission of clusters out to the virial radius, especially for higher-redshift cool-core clusters, for which there might be a non-negligible contribution from the bright cluster core to the emission in the outer volumes and from secondary scatter by sources outside the field of view. Note that the properties of the particle background are well known via the stowed background observations, with uncertainties in the background across its spectrum which are within $\lesssim 2$ percent (\cite{hickox2006} and see also discussion in \cite{sun2009}).

This data analysis project represents a follow-up of our previous works on studying cluster thermodynamic properties, including emission measure, gas density, and gas fraction, out to $R_{200}$ \citep{morandi2013b,morandi2014}. Indeed, when data of sufficient depth are available, we have shown that {\em Chandra} can accurately measure the surface brightness and temperature of clusters out to $R_{200}$ \citep[e.g., in Abell~1835 and Abell~133]{bonamente2013,morandi2013b,morandi2014}. The {\em Chandra}'s superior angular resolution enables robust identification and removal of point sources from the X-ray images, while minimizing the contribution from the bright cluster core to the emission in the outer volumes -- both challenging tasks for {\em Suzaku} --  and to provide an independent confirmation of the {\em Suzaku} results. Observational constraints on the gas clumping factor inferred from the inhomogeneities of the X-ray surface brightness are also significantly smaller than those inferred from the {\em Suzaku} measurements \citep[$C\sim1.5-2$ at $R_{200}$,][]{eckert2013a,morandi2013b,morandi2014}; the gas density slope reaches $\beta \sim 0.9$ at $R_{200}$, in good agreement with the predictions of hydrodynamical simulations, but steeper than the values inferred from recent {\em Suzaku} observations \citep{bautz2009,kawaharada2010}. 

At present, {\em Suzaku} has mainly studied high mass clusters to the virial radius, primarily at low redshift, given its large PSF ($\sim 2'$), which limits the size of the annuli used in the spatially resolved spectral analysis, reducing the angular resolution of the temperature and entropy profiles \citep{bautz2009,urban2013}. It is clearly important to study clusters at higher redshift to investigate e.g. the evolution of the thermal properties of the ICM, and with independent X-ray observatories. 

In order to study the physical properties of the ICM in the virialization region we exploit the large archival dataset of {\em Chandra} clusters. We study the surface brightness profiles of a sample of 320 clusters ($0.056 <z< 1.24$) with $kT\gesssim 3$ keV, observed with {\em Chandra}. We stacked\footnote{We caution the reader that, from now on, with the word ``stacking'' we do not refer to an actual co-adding of the images, but rather to the median of the distribution of the radial emission measure profiles $EM$ at radii $R/R_{200}$, once re-scaled according the self-similar model (see \S\ref{selfsim} for further details).} the emission measure profiles $EM\propto \int n_e^2\, dl$ of the cluster sample to detect a signal out to and beyond $R_{200}$. We then measured the average emission measure, gas density and scatter in cluster outskirts. This pioneering work for the {\em Chandra} archive follows the original idea successfully applied in {\em ROSAT} (\citealt{eckert2012}, see also \citealt{dai2007}); and with the Sunyaev Zeldovich maps \citep{plagge2010,planck2013}, which allowed to detect the physical properties beyond $R_{200}$. 

The paper is organized as follows. In \S\ref{satecn} we present the cluster sample, in \S\ref{dataan} we describe the X-ray analysis and in \S\ref{sys453455y4} we discuss the systematics in the data analysis. The results on the gas density and gas mass fraction are presented in \S\ref{sys453b} and \S\ref{syeebse2}. In \S\ref{conclusion33} we summarize our conclusions. Throughout this work we assume the flat $\Lambda$CDM model, with matter density parameter $\Omega_{\rm m}=0.3$, cosmological constant density parameter $\Omega_\Lambda=0.7$, and Hubble constant $H_{0}=100h \,{\rm km\; s^{-1}\; Mpc^{-1}}$ where $h=0.7$. Unless otherwise stated, we report the errors at the 68.3\% confidence level.

\begin{table*}
\begin{center}
\caption{Table \ref{tab:1} is published in its entirety in the electronic edition of the {\em Monthly Notices of the Royal Astronomical Society}. A portion is shown here for guidance regarding its form and content. In the present table we report the X-ray properties of the galaxy clusters in our sample. For each object different columns report the cluster name, the redshift $z$, the exposure time $t_{\rm exp}$,  the identification number of the {\it Chandra} observations, the spectroscopic temperature $T_{\rm ew}$ in the radial range $0.15-0.75\; R_{500}$, the centroid shift (in units of $R_{500}$), and a flag for the presence or not of a cooling core (labeled CC and NCC, respectively). }
\begin{tabular}{l@{\hspace{0.8em}} c@{\hspace{0.8em}} c@{\hspace{0.8em}} c@{\hspace{0.8em}} c@{\hspace{0.8em}} c@{\hspace{0.8em}} c@{\hspace{0.8em}} c@{\hspace{0.8em}} c@{\hspace{0.8em}} c@{\hspace{0.8em}} c@{\hspace{0.8em}} c@{\hspace{0.8em}} c@{\hspace{0.8em}} }
\hline \\
Cluster   & $ z $ & $ t_{\rm exp} $ & ID &   $T_{\rm ew}$ & $w$ & CC/NCC\\
          &        &        ks         & &  (keV)        & ($\times 10^{-2}$)  &\\
\hline \\
             A85&   0.056 &   272.4&    904  4881  4882  4883  4884  4885  4886  4887& $   6.45\pm   0.86$ & $   0.49\pm   0.18$ &     CC \\
 &  & &    4888 15173 16264 15174 16263&  &\\
              A133&   0.057 &  2393.0&  13442 13443 13444 13445 13446 13447 13448 13449& $   3.76\pm   0.25$ & $   0.26\pm   0.09$ &     CC \\
 &  & &   13450 13451 13452 13453 13454 13455 13456 13457&  &\\
 &  & &   14333 14338 14343 14345 14346 14347 14354 13391&  &\\
 &  & &   13392 13518  3183  3710 12177 12178 12179  9897&  &\\
              A644&   0.070 &    48.9&   2211 10420 10421 10422 10423& $   8.12\pm   1.12$ & $   1.56\pm   0.57$ &     CC \\
              A401&   0.075 &   153.9&  10416 10417 10418 10419 14024& $   5.44\pm   0.86$ & $   1.03\pm   0.38$ &    NCC \\
             A2029&   0.076 &    29.2&   6101 10434 10435 10436 10437& $   7.38\pm   0.92$ & $   0.21\pm   0.08$ &     CC \\
             A1650&   0.084 &   218.4&   7691  5822  5823  6356  6357  6358  7242 10424& $   5.89\pm   0.86$ & $   0.38\pm   0.14$ &     CC \\
 &  & &   10425 10426 10427&  &\\
             A1068&   0.137 &    19.5&  13595 13596 13597 13598& $   6.38\pm   1.26$ & $   0.29\pm   0.11$ &     CC \\
             A2276&   0.141 &    39.5&  10411& $   3.40\pm   0.48$ & $   0.49\pm   0.18$ &    NCC \\
             A1413&   0.143 &   155.6&   5003 12194 12195 12196 13128  1661  5002   537& $   7.78\pm   0.22$ & $   0.57\pm   0.21$ &     CC \\
 &  & &    7696&  &\\
             A3402&   0.146 &    19.8&  12267& $   3.09\pm   0.55$ & $   1.39\pm   0.51$ &    NCC \\
  \hline \\
\end{tabular}
 
\label{tab:1}
\end{center}
\end{table*}

\section{Cluster sample}\label{satecn}
\begin{figure}
\begin{center}
\psfig{figure=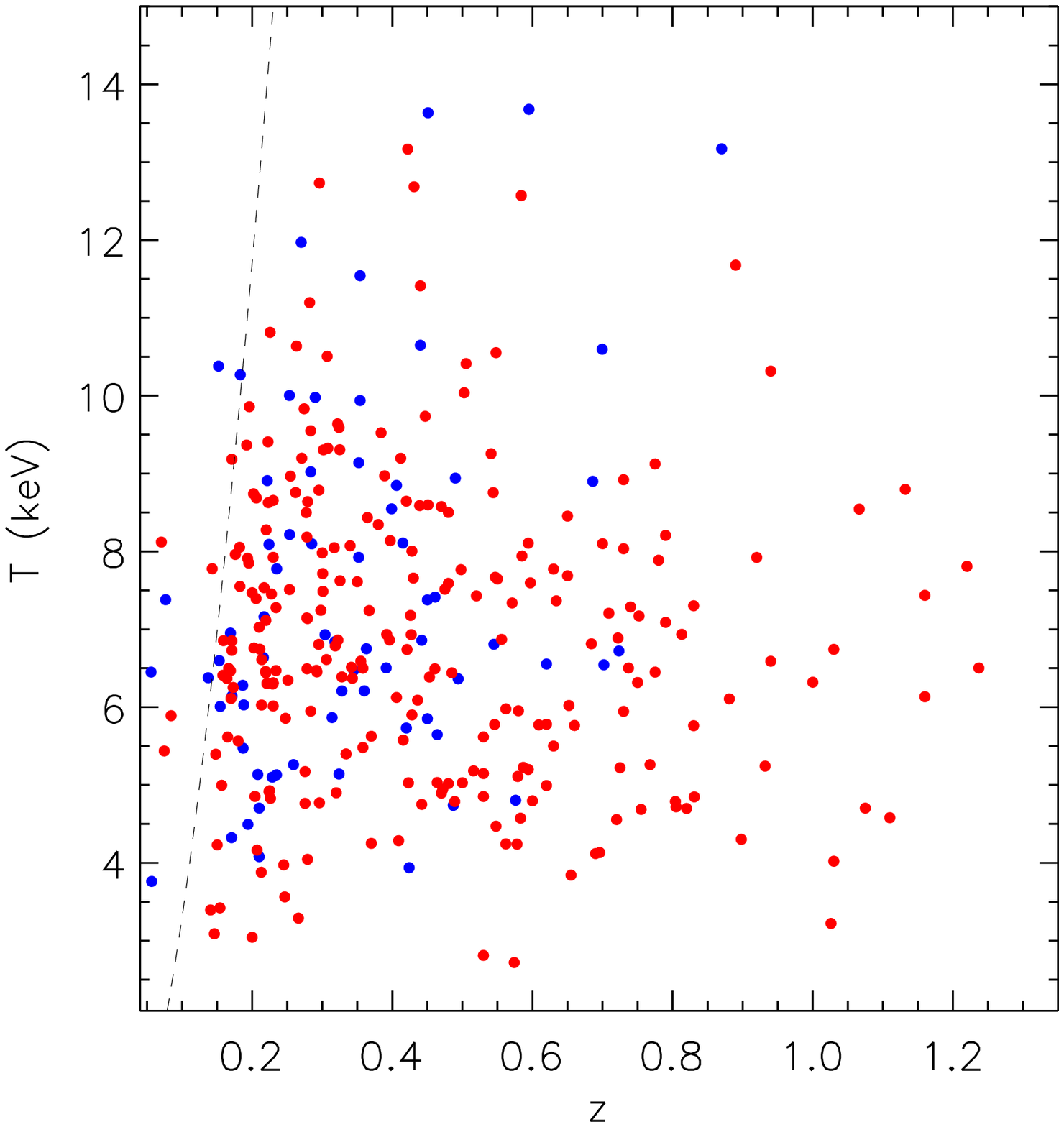,width=0.44\textwidth}
\caption{Distribution of the clusters in our sample in a temperature versus redshift diagram. The dashed line marks the region of the plot corresponding to 10 arcmin (the ACIS--I {\em Chandra} fov is $16\times 16$ arcmin). Note the some clusters, e.g. A133, A644, A2029, A1650 and A85, at low redshift ($\sim0.05$), are characterized by values of $R_{200}$ larger than the fov of {\em Chandra}. These clusters have a large number of mosaic observations covering the outskirts. The blue and red dots correspond to clusters classified as cool core and non-cool core, respectively (see \S\ref{morph}).}
\label{T-z}
\end{center}
\end{figure}

The sample we chose for this analysis is composed of 320 clusters from the {\em Chandra} archive (Table \ref{tab:1}), with redshift range $z=0.056-1.24$ (the sample median redshift is $\sim0.4$) and temperatures $kT\gesssim 3$ keV (the sample median temperature is $\sim 7$ keV). These sources have been observed with the ACIS-I imaging array. These clusters have been selected because the X-ray observations encompass $R_{200}$ of each cluster, hence they are suitable for observations of cluster outskirts. Moreover, source-free regions of the cluster observation are always present, allowing measurements of the local background (Figure \ref{T-z}). The modeling of the background is indeed crucial for robust measurements of the physical parameters in the outer regions.

This sample encompasses most of the clusters from large samples studied in previous works, including e.g.  \cite{vikhlinin2006}, \cite{maughan2008}, \cite{cavagnolo2009}, \cite{ettori2009b} and \cite{McDonald2013}.

Although the sample was selected based on the quality of the existing observations and hence might be subject to selection effects, for the purpose of this work we did not require that the sample be complete, given the high level of self-similarity of the cluster outskirts. The X-ray properties of the galaxy clusters in our sample are presented in Table \ref{tab:1}, while the number of net counts per bin for each cluster and the distribution of total net X-ray counts are shown in Figure \ref{sample4634f}.

\section{The self-similar model: measuring the physical parameters}\label{selfsim}

Stacking the emission measure of a cluster population is possible thanks to the high level of self-similarity of the cluster outskirts.

In this respect, the self-similar model \citep[see, e.g.,][]{kaiser1986} gives a simple picture of the process of cluster formation in which the ICM physics is driven by the infall of cosmic baryons into the gravitational potential of the cluster DM halo. Power law scaling relations are expected under simplified models in which clusters are self-similar objects, having formed in single monolithic gravitational collapses and whose ICM is heated only by the shocks associated with the collapse. The underlying idea is that gravity is the only responsible for the observed values of the different physical properties of galaxy clusters; hence clusters are identical objects when scaled by their mass or temperature \citep{lau2014}.

Numerical simulations confirm that the DM component in clusters of galaxies, which represents the dominant fraction of the mass, has a remarkably self-similar behavior; however the baryonic component does not show the same level of self-similarity. For instance, deviation of the $L-T$ relation in clusters with respect to the theoretical value predicted by the previous scenario suggest that some energetic mechanism, in addition to gravity, such as (pre)-heating and cooling \citep{Borgani2005,bryan2000,morandi2007b,sun2012} intervene to break the expected self-similarity of the ICM in the innermost regions. Consequently, the comparison of the self-similar scaling relations to observations allows us to evaluate the importance of the effects of non-gravitational processes on the ICM physics. While the central regions of clusters often exhibit complicated physical phenomena, such as AGN heating and cooling flows, the underlying physics of the outskirts is relatively simple, gravity-dominated and hence nearly self-similar. 

Assuming the spherical collapse model for the DM halo and the equation of hydrostatic equilibrium to describe the distribution of baryons into the DM potential well, in the self-similar model the cluster mass at an overdensity $\Delta$ (e.g. $\Delta=500,200$) and temperature are related by:
\begin{equation}
E_{z} \Delta_z^{1/2} M_{\rm tot} \propto T^{3/2}\ ;
\end{equation}
where the factor $\Delta_z= \Delta \times \left[1 +82 \left(\Omega_z-1\right) / \left(18 \pi^2\right) -39 \left(\Omega_z-1\right)^2 / \left(18 \pi^2\right) \right]$, with $\Omega_z = \Omega_{\rm 0m} (1+z)^3/E_z^2$, accounts for evolution of clusters in an adiabatic scenario \citep{kaiser1986}. So we have $ R_{\Delta_z} \propto {\left({M/(\rho_{c,z}\Delta_z)}\right)}^{1/3} \propto T^{1/2}E_{z}^{-1}\Delta_z^{-1/2}$.  By setting $f_z\equiv E_{z}{(\Delta_z/\Delta)}^{1/2}$ \citep[e.g.][]{ettori2004}, from the previous equations we can easily obtain the following relations:
\begin{eqnarray}
f_z R_{\Delta_z} \propto T_{\rm gas}^{1/2}\nonumber \\
f_z M_{\rm gas} \propto  T_{\rm gas}^{3/2}
\end{eqnarray}
From the previous equation it follows that:
\begin{equation}
EM \propto \int n_e^2\; dl \propto M_{\rm gas}^2/R_{\Delta_z}^5 \propto  f_{\rm gas}^2 \; f_z^3 T_{\rm gas}^{1/2}
\label{dweded}
\end{equation}
assuming that all clusters have the same gas mass fraction $f_{\rm gas}$. Note that our sample is characterized by a relatively narrow temperature range (larger than 3 keV), such that the physical parameters of the clusters in our sample should show little dependence on gas fraction. The electron density $n_e(r)$ can be finally inferred by deprojecting the stacked X-ray emission measure profile (Equation \ref{dweded}) via the spherical onion peeling method \citep{morandi2013a}. 

In our work we have considered all the physical quantity at fixed overdensity ($\Delta_z=\Delta$), i.e. $f_z=E_{z}$ in the above equations.

In order to infer $R_{500}$ from the observables quantities, we rely on the existence of low-scatter scaling relations between the total thermal energy $Y_X=M_{\rm gas}\, T_{X}$ and the total mass, as predicted by self-similar theory and confirmed by high-resolution cosmological simulations \citep{kravtsov2006}. In this respect we used the $Y_{X}-M$ relation measured for the \citet{sun2009} sample of clusters:
\begin{equation}\label{e.ym}
E_{z}^{2/5} M_{500}  =  \left(\frac{h}{0.7}\right)^{\frac{5B}{2}-1} \!\!\!\!\!\!\!\!\!\!\! 1.63\times10^{14} {\left(\frac{Y_X}{3.60\times10^{13} {\rm M_{\odot} keV}}\right)}^{0.571}
\end{equation}
with $B=0.571$. Hence we recovered $R_{500}$. 

We point out that we determined $T_X$ in the radial range $0.15-0.75\; R_{500}$ (\S\ref{laoa3}), while the previous relation has been calibrated in the range $0.15-1\; R_{500}$. Our choice is motivated by the fact that this is a straightforward measurement for our intermediate and high-z clusters because exposures were designed to provide a sufficient statistical accuracy to infer the spectral temperature in the aforementioned radial range. We correct our $T_X$ measurements in order to match the definition employed in Equation \ref{e.ym} via mock simulations by assuming our average gas density profile (\S\ref{sys453b}) and the average temperature profile as measured by \cite{vikhlinin2006}. This correction is $0.96\pm0.02$.

$R_{500}$ and $Y_X$ were computed iteratively, measuring the temperature in the aperture $0.15-0.75\; R_{500}$ and the gas mass within $R_{500}$, computing a new ${Y_X}$, and hence estimating a new value of $R_{500}$.

In order to infer $R_{\Delta}$ ($\Delta=500,200,100$) from the known value of $R_{500}$, we apply the following statistical method. First, we assumed a NFW distribution with parameters $(c,R_{200})$; second,  we inferred a value of the concentration parameter via the $c-M_{\rm vir}$ relation of \cite{duffy2008} based on the results from numerical simulations. This provides an estimate of $R_{\Delta}$ for given value of $R_{500}$. For our sample we have $R_{500}=(0.65\pm0.01) R_{200}$ and $R_{100}=(1.36\pm0.01) R_{200}$

The gas fraction within an overdensity $\Delta$ can be computed directly from the gas density profiles:
\begin{equation}
f_{gas,\Delta}=\frac{M_{gas,\Delta}}{M_\Delta} =\frac{3}{\Delta\rho_{c,z}}\int_0^1\rho_{gas}(x)x^2\, dx.
\label{fgas235}
\end{equation}
with $x={R}/{R_{\Delta}}$.

\section{X-ray analysis}\label{dataan}
Description of the X-ray analysis methodology can be found in \cite{morandi2013b}. Here we briefly summarize the most relevant aspects of our data reduction and analysis.

\subsection{X-ray data reduction}\label{laoa}
All data were reprocessed from the level 1 event files using the \textit{CIAO} data analysis package -- version 4.6.1 -- and the calibration database (CALDB 4.6.3) distributed by the {\it Chandra} X-ray Observatory Center. Note that the ACIS QE contamination model, which has been released in CALDB and is associated with the deposition of one or more materials on the ACIS detectors or optical blocking filters, is the latest available (version N0009).

We analyzed 613 datasets, corresponding to 320 galaxy clusters, retrieved from the NASA HEASARC archive with a total exposure time $\sim 20$~Ms for the whole sample. All the observations are carried out by using the ACIS--I CCD.

We reprocessed the level-1 event files to include the appropriate gain maps and calibration products. As part of the data reduction, corrections were made for afterglows, charge transfer inefficiency (CTI), bad pixels and solar flares. We used the \texttt{acis\_process\_events} tool to check for the presence of cosmic-ray background events, correct for spatial gain variations due to charge transfer inefficiency and re-compute the event grades. Then we filtered the data to include the standard events grades 0, 2, 3, 4 and 6 only, and therefore we filtered for the Good Time Intervals (GTIs) supplied, which are contained in the {\tt flt1.fits} file. 

A careful screening of the background light curve is necessary to discard contaminating flare events. A common way of removing these periods of particle background flares is to create a lightcurve of the local background, once point sources of high and variable emission are excluded. The lightcurve is then created by using the tool {\tt dmextract}. In order to clean the datasets of periods of anomalous background rates, we used the {\tt deflare} script, so as to filter out the times where the background count rate exceed $\pm 3\sigma$ of the mean value. Most observations were taken in VFAINT mode, and in this case we applied VFAINT cleaning to both the cluster and blank-sky observations. Finally, we filtered the ACIS event files on energy selecting the range 0.3-12 keV, so as to obtain a level-2 event file. 

Point sources and extended substructures were detected and removed using the script \texttt{wavedetect}, which provides candidate point sources, and the result was then checked through visual inspection.

\subsection{X-ray surface brightness analysis}\label{laoa3}
We produced X-ray images from the level-2 event file. Our goal is indeed to measure the emission measure and the gas density profile in a non-parametric way from the surface brightness. Multiple observations were reduced individually to apply the correct calibration to each dataset. The cluster surface brightness profile is obtained from merged images.

The X-ray images were extracted from the level-2 event files in the energy range $0.7-2.0$ keV. This energy band was chosen in order to: i) maximize the S/N in the outskirts of the clusters, and ii) minimize the dependency of the cooling function on the temperature and metallicity. Concerning the latter point, the integrated cooling function in the $(0.7-2)$~keV band is approximately given by $\Lambda(T) \propto T^{-\alpha}$, with $-0.02\la \alpha\la 0.2$ for $T \gesssim 2$ keV (in all the clusters the temperature is not expected to fall below $\sim$2 keV), such that systematic uncertainties in the estimated projected temperature have a negligible impact on the gas density. In particular, we point out that we inferred the density profiles out to $R_{100}$ from the stacked emission measure profiles.  

We then corrected the images by the exposure maps to remove the vignetting effects. We created an exposure-corrected image from a set of observations using the \texttt{merge\_obs} tool to combine the ACIS--I observations. All maps were checked by visual inspection at each stage of the process.

Next, we determined the centroid ($x_{\rm c},y_{\rm c}$) of the surface brightness image by locating the position where the derivatives of the surface brightness variation along two orthogonal (e.g., X and Y) directions become zero. 

Since for all the clusters the field-of-view encompasses the X-ray boundary ($R_{100}$), the emission does not extend across the entire detector, and we have a sufficiently large region to obtain a local background. In all the clusters the local background was measured from a region beyond $1.5\,R_{200}$, where the surface brightness was approximately constant, i.e. it has reached the background level. 

The local background is a combination of a particle component that is not vignetted, and a sky component that is vignetted. To determine the surface brightness of the cluster and of the local soft X-ray background, an accurate procedure consists of subtracting the non-vignetted particle component as measured from {\em Chandra} observations in which the ACIS detector was stowed, after rescaling the stowed background to match the $9.5-12$ keV cluster count rate of the observations (where the {\em Chandra} effective area is negligible and thus very little cluster emission is expected). Indeed, although the particle background flux may vary with time, \cite{hickox2006} has shown that the spectral distribution of the particle background is remarkably stable, even in the presence of changes in the overall flux, and that the ratio of soft-to-hard (2-7 keV to 9.5-12 keV) count rates remains constant to within $\lesssim 2$ percent in time. We verified that all our sources are characterized by remarkably stable background by comparing the spectra of the local background and of the blank sky fields. 

In order to correct for vignetting, a common approach is to divide the counts image by an exposure map, weighted according to a specific model for the incident spectrum, to rescale all parts of the image to the same relative exposure. Nevertheless, since the exposure map is both energy and position dependent, we cannot assume the same exposure map for source and background emission. Our model of the desired source surface brightness $S_{X,s}$ breaks down the observed emission  $S_{X,obs}$ into four components:
\begin{eqnarray}
S_{X,obs}({\bf x}) \propto & S_{X,s}({\bf x})\cdot C_s({\bf x},E) +S_{X,bkg}({\bf x})\cdot C_{bkg}({\bf x},E)  \nonumber\\
& + S_{X,pb}({\bf x},E(t)) + S_{X,r}({\bf x},E)
\label{1.em.x.eq22}
\end{eqnarray}
where $S_{X,s}({\bf x})$ is the source surface brightness in the pixel ${\bf x}=(x,y)$, $S_{X,bkg}$ is the cosmic X-ray background (CXB) background and $S_{X,pb}$ is the (time dependent) particle background. $C_s({\bf x},E)$ and $C_{bkg}({\bf x},E)$ refer to the exposure maps for source and background, respectively. 

$S_{X,r}({\bf x},E)$ refers to the readout artifact background. We simulated, re-normalized and accounted for this readout background using the {\tt make\_readout\_bg} routine (see \cite{hickox2006} for further details).

The radial surface brightness profile is thus derived with the exposure correction and particle background subtraction using the scaled stowed background. The region where the CXB is more dominant than the cluster emission can be determined from the flattened portion at the outer region of the surface brightness profile. We applied a direct subtraction of the CXB+particle background by means of Equation \ref{1.em.x.eq22}. In order to calculate $C_{bkg}({\bf x},E)$, we modeled the soft CXB component by an absorbed power law with index 1.4 and two thermal components at zero redshift, one unabsorbed component with a fixed temperature of 0.1 keV and another absorbed component with a temperature derived from spectral fits \citep[$\sim0.25$ keV,][]{sun2009}. 

The boundary radius of the X-ray and surface brightness analysis corresponds to a ratio of source to background flux $\sim$ 15-40 percent at $R_{200}$. Thus, a careful analysis of the systematics and how these uncertainties propagate into the determination of the physical parameters is required (\S\ref{sys453}).

With robust background subtraction and modeling, from the X-ray images we measure the emission measure profile $EM\propto \int n_e^2\, dl$ and calculated the median distribution to compute stacked profiles. We then scaled the emission measure profiles and the deprojected density profiles for the clusters in our sample. A self-similar scaling was applied to the emission-measure profiles, i.e. each profile was rescaled by the quantity $E_z^3({kT}/{10\mbox{ keV}})^{1/2}$ (\S\ref{selfsim}), following their evolution with redshift and dependency on the cluster mass, since the profiles is expected to show a remarkable level of self-similarity outside of the core ($R > 0.2 R_{200}$). From the deprojection of the median emission measure profile we then recover the expected density profile.

\subsection{X-ray spectral analysis}\label{laoa34}
The global temperatures have been recovered by fitting spectra from different observations simultaneously. The spectral analysis was performed by extracting the source spectra from circular annuli around the centroid of the surface brightness and by using the {\em CIAO} \texttt{specextract} tool from each observation. The spectral fit was performed by simultaneously fitting an absorbed thermal emission model in the energy range 0.7-7 keV and in the region $0.15-0.75\; R_{500}$. We used the XSPEC package \citep[][version 12.8.2]{1996ASPC..101...17A} to perform the spectral fit. We adopted the APEC emissivity model \citep{foster2012} and the AtomDB (version 2.0.2) database of atomic data, and we employed the solar abundance ratios from \cite{asplund2009}. We also used the Tuebingen-Boulder absorption model (\texttt{tbabs}) for X-ray absorption by the ISM. Free parameters in the APEC model are temperature, metal abundance, normalization. We fixed the hydrogen column density $N_H$ to the Galactic value by using the Leiden/Argentine/Bonn (LAB) HI-survey \citep{kalberla2005}. The redshift to the value obtained from optical spectroscopy. We also group photons into bins of at least 20 counts per energy channel and applying the $\chi^2$-statistics. We analyzed the observations individually to check for consistency before jointly fitting multiple datasets referring to a single cluster. The background spectra have been extracted from regions of the same exposure for the ACIS--I observations, for which we always have some areas ($\gesssim R_{100}$) free from source emission. 

We also checked for systematic errors due to possible source contamination of the background regions, which has been estimated by considering the predictions of the surface brightness from hydrodynamical numerical simulations including cooling, star formation and supernovae feedback \citep{roncarelli2006}. The contamination is always of the order a few percent, and this translates into a negligible bias of the determination of the spectral global temperature. 

We finally validated the method to recover the global temperature for each cluster by creating a mock spectrum via a single-temperature absorbed thermal model with parameters fixed to those from the aforementioned spectral analysis, and then adding a background spectrum from a region free of the source emission. We thus repeated the spectral analysis on the mock datasets and we found a very good agreement between input and recovered spectral parameters.

In the appendix we present a comparison between spectral temperatures and metallicities recovered via the AtomDB databases 2.0.2 and 1.3.1, via different models of X-ray absorption by the ISM, and by accounting also for the molecular and ionized hydrogen in the hydrogen column densities.

\subsection{Morphological type classification}\label{morph}
In this section we describe how the clusters were classified based on their morphological type, i.e. based on the presence of a cool core and/or dynamical state of clusters. Dividing the data into these subsamples allows us to investigate the effect of cool cores and morphological disturbance on a cluster’s position with respect to the mean relation.

We adopted the centroid shift method as our method to infer to the dynamical state of clusters \citep{maughan2008}. Centroid shifts have been measured from the variation of the centroid of cluster emission within apertures of increasing radii $R_i$, which have similarly been previously used to infer the emission measure. The centroid shift, $ w $, was defined as the standard deviation of these centroids in units of $R_{500}$. 

Centroid shifts were measured from exposure-corrected images in order to eliminate the effects of vignetting. We masked out point sources, chip gaps and all the substructures which have been similarly excluded in recovering the emission measure profile. This generates a 2D mask $I(x-x_{\rm c},y-y_{\rm c})$, with $x_{\rm c},y_{\rm c}$ being the centroid as determined in \S\ref{laoa3}. We then symmetrized this mask, i.e. $I(x-x_{\rm c},y-y_{\rm c})=I(-x+x_{\rm c},-y+y_{\rm c})$, in order to remove any spurious dependency of the centroid shifts on the adopted masking, i.e. a perfect azimuthally symmetric source must generate a zero centroid shift regardless of the adopted mask. This definition adds the benefits to robustly relate $w$ to the global dynamical state of clusters, while eliminating the impact of point sources, chip gaps and substructures similarity masked in the emission measure profiles.

The measured centroid shifts are summarized in Table \ref{tab:1}. The errors quoted for the $w$ are the statistical uncertainties on a standard deviation calculated from $n$ measurements ($w\sqrt{2/(n-1)}$). Objects with center shift parameter $w > 0.01 R_{500}$ are classified as morphologically disturbed. 

Next, systems are classified as cool cores (CC) or non-cool cores (NCC) on the basis of density $E_z^{-2} n_{e,0}$ at $0.03\,R_{500}$. The threshold we use to define a cool core systems is $E_z^{-2} n_{e,0}>1.5\times 10^{-2}{\mbox{cm}}^{-3}$. We compared our definition of CC systems with \cite{cavagnolo2009} in \S\ref{maugh1ehrhgr}, whose sample shares 93 objects in common with ours. They use the entropy threshold $K_0=30-50 \, \mbox{keV cm}^{2}$  to approximately demarcate the division between CC and NCC. 

\begin{figure}
\begin{center}
\psfig{figure=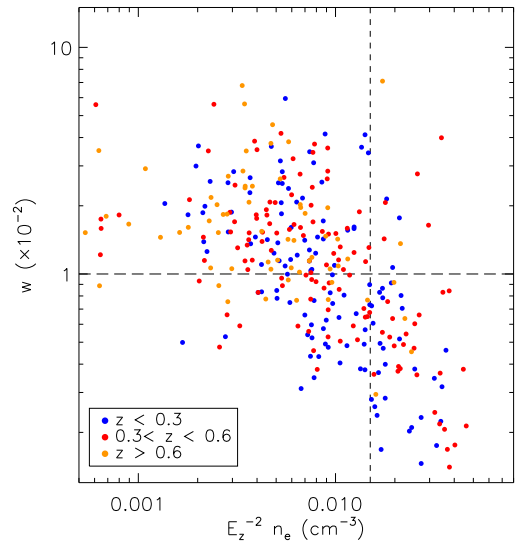,width=0.44\textwidth}
\caption{Plot of centroid shifts versus electron density at $0.03\,R_{500}$ for the whole sample.}
\label{w-cc}
\end{center}
\end{figure}

Note that centroid shifts and central density are broadly anti-correlated (Figure \ref{w-cc}), being both a proxy of the dynamical state of a cluster, i.e. cool-core system have larger central density and overall more relaxed dynamical state.  

We finally measured the flattening and orientation of the X-ray surface brightness. We computed the moments of the surface brightness within a circular region of radius $R_{500}$ centered on the centroid of the X-ray image (see \cite{morandi2010a} for further details on this method). This allows us to estimate the ICM eccentricity on the plane of the sky and the orientation (position angle) of an elliptical X-ray surface brightness distribution. We use this position angle as a proxy of the direction of the large-scale filament (see discussion in \S\ref{lsst3}). 

\section{Analysis of the systematics in the data analysis}\label{sys453455y4}

\subsection{Validation of the stacking method}\label{sggg53}
We validated our stacking method via simulations of mock datasets including systematics. From the rescaled emission measure, averaged on the whole sample, we created mock azimuthally-averaged source brightness images $S_{X,s}({r})$ for each observation, according to the observation aspect information. We then used blank-sky fields to produce realistic backgrounds $S_{X}^{bkg}({\bf x})$ for each observation. We extracted background images from the blank-field background data provided by the ACIS calibration team in the same chip regions as in the observed cluster. The blank-sky observations event files underwent a reduction procedure consistent with that applied to the cluster data, after being reprojected onto the sky according to the observation aspect information by using the {\tt reproject\_events} tool. We then added these background images to the source brightness. In order to estimate the contribution of the particle background, we used stowed background observations, renormalized to match the blank-sky count rate at high energy (9.5-12 keV). We point out that in this way we capture biases due both possible spatial variation of the CXB and temporal variations of the particle background spectrum.  We finally applied the whole procedure described in \S\ref{laoa} (Equation \ref{1.em.x.eq22}) on the mock observations. A comparison between the true and measured emission measure profile reveals that the distribution of the residual (normalized by the measurement uncertainties) follows the expected Gaussian distribution with zero mean, with reduced $\chi^2$ of the order of the unity. This indicates that the measurement errors fairly represent the error budget and our recovered physical parameters are not significantly biased.

\subsection{Systematics in background modelling}\label{sys453}

Accurate measurements of the ICM in the cluster outskirts require an accurate knowledge of the components of the X-ray background, and an accurate understanding of how the uncertainties of the background modeling propagate on the desired physical parameters.

For more details and a thorough validation of the method to gauge statistical and systematic errors, we refer the reader to \cite{morandi2014}. Here we briefly outline the key aspects.

The cosmic X-ray background (CXB) consists of the local hot bubble emission, the thermal emission of the galactic halo and the contribution of unresolved point sources (mostly AGNs). The latter component is modeled as a power law with index 1.4. When analyzing regions of finite size we must account for Poisson fluctuations of these unresolved point sources. The expected deviation from the average value for a given observed solid angle $\Omega$ resolved to a threshold flux $S_{\rm thres}$ is:
\begin{eqnarray}
\sigma^2_{\rm CXB} = (1/\Omega) \int_{0}^{S_{\rm thres}} \Big(\frac{dN}{dS}\Big)\times S^2 ~ dS
\label{flux56}
\end{eqnarray} 
where ${dN}/{dS}$ is the cumulative flux distribution of point sources as employed by \cite{moretti2003}. The threshold flux $S_{\rm thres}$ has been calculated by determination of the local flux limit to which we can robustly identify a point source, commonly known as sensitivity map \citep[see, e.g.,][for further details on the method]{ehlert2013}. Our joint {\em Chandra} observations of the outskirts allow the CXB to be resolved to a threshold flux $S_{\rm thres} \sim 10^{-15}$ erg cm$^{-2}$ s$^{-1}$ deg$^{-2}$ in the soft band. Solving Equation \ref{flux56}, this translates into an error in the range $\lesssim$3\% on the surface brightness in the outer volumes for the average emission measure, while it becomes negligible in the inner volumes. If we consider the subsample of clusters at higher redshift ($z>0.3$), this error becomes appreciable ($\lesssim 10$ percent in the outer volumes).

We point out that the previous analysis does not take into account the possible systematics due to variation of the PSF across the ACIS-I detector, the PSF being about one order to magnitude larger ($\sim 10-20$ arcsec) at the edges of the CCD with respect to the aimpoints ($\sim 1$ arcsec); neither it accounts for the different contamination of the galaxy cluster signal in detecting point sources across the CCD. Indeed, the signal of point sources near the edge of the field would be diluted on larger angular scales due to the larger {\em Chandra} PSF, but at the same time would be less contaminated by the cluster signal, the cluster center being observed customarily near the center of the CCD. The first (second) effect would make more difficult (easier) to detect point sources with respect to the CCD center via the wavelet decomposition algorithm described in \S\ref{laoa}. In order to gauge the impact of these systematics we perform Monte Carlo (MC) simulations of the point sources as employed by \cite{moretti2003}, which should account for most of the CXB background. We then added particle and galactic background, which has been convolved with the {\em Chandra} PSF at the position $(x,y)$ of the CCD. We finally added mock cluster signal (see \S\ref{selfsim} for details on the mocks). For each MC simulation we extracted point source by mimicking the procedure of wavelet detection algorithm. This allow us to estimate the bias (downwards) due to the aforementioned systematics, which is $\lesssim 3\%$(4\%) on the gas density at $R_{200}$($R_{100}$).

We then estimated further systematics in the data analysis due to background subtraction. For this purpose, we use stowed background and local background taken from source-free regions of the cluster observations in order to model the contribution of the CXB and particle background, respectively. The two major sources of systematic uncertainties are: i) the renormalizion of the stowed background to match the cluster count rate at high energy (9.5-12 keV) due to Poisson errors; and ii) uncertainties in the particle background across its spectrum. Note that the properties of the particle background are well known via the stowed background observations, with uncertainties in the background across its spectrum which are within $\lesssim 2$ percent (\cite{hickox2006} and see also discussion in \cite{sun2009}). 

Finally, we estimated the uncertainty in our measurements due to soft X-ray background variations across the field of view. We divided the regions used to obtain the background model into independent subregions with different azimuthal direction with respect to the cluster center. We implemented a bootstrap approach, where we randomly picked these background subregions, allowing repetitions, and re-determined the background level via Equation \ref{1.em.x.eq22}. We repeated this procedure $10^5$ times, and determined the distribution of the average background. The uncertainties due to CXB fluctuations are always smaller than the statistical error bars determined by assuming that the soft X-ray background is spatially constant.

All the aforementioned uncertainties on the background modelling (cosmic X-ray background, particle background, soft X-ray background variations and unresolved point sources) have been propagated in recovering the desired physical parameters, by means of MC randomizations of the background including both statistical and systematic uncertainties. 

\subsection{The impact of undetected clumps, gas inhomogeneities and multi-temperature distribution}\label{bias11}

The CDM scenario predicts a picture where clumps at subvirial temperature are infalling along filaments and accreting onto the cluster outskirts. Given the low signal-to-noise, low sensitivity/integration time and/or poor spatial resolution of the X-ray telescopes, it might be impossible to mask out cold substructures e.g. in the X-ray analysis, and they might remain undetected. A cold phase ($T\lesssim 1$ keV) of the ICM would also coexist in the outskirts with the hot ICM at virial temperatures ($\gesssim 2-4$ keV), leading to a multiphase structure of the ICM. The presence of a multiphase structure of the ICM will bias the spectroscopic temperatures and gas density. Spectral temperature measurements indeed customarily hinge on a single-temperature absorbed thermal model, since the data quality is in general not sufficient to constrain a multitemperature model, especially in the outskirts where the signal is overwhelmed by the noise.

\begin{figure}
\begin{center}
\psfig{figure=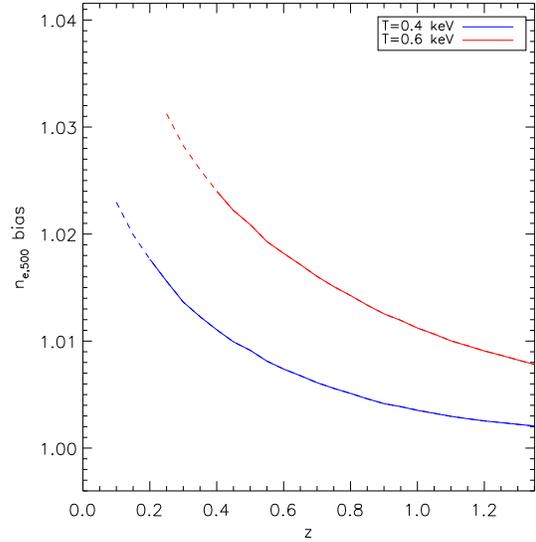,width=0.44\textwidth}
\caption{Bias (upwards) on the gas density at $R_{500}$ due to undetected substructures at subvirial temperatures T$\sim0.4$ keV and 0.6 keV. The plotted redshift corresponds to the range where the subclumps fall below the detection threshold of the observation (\S\ref{sys453}), i.e. they are undetected. We assumed a background level of $2\times10^{-11}$ erg cm$^{-2}$ s$^{-1}$ deg$^{-2}$ in the soft band, a filling factor of 1\% for these subclumps, a cluster temperature of 4 keV and self-similar gas density profile for both cluster and subclumps. The solid (dashed) line refers to an observation time of 100 (10) ks.}
\label{bias}
\end{center}
\end{figure}

The presence of undetected dense and cold clumps at temperatures $0.5\lesssim kT \lesssim 2$ keV in the ICM will typically enhance the X-ray emission, since the instrument is very sensitive to the low energy features of these clumps whose exponential bremsstrahlung cutoff will be at energies $\sim kT/(1+z)$. Therefore, besides upon the dynamical state of the cluster, the impact of dense clumps will also depend on the redshift of the observation, the {\em Chandra} effective area being small for photon energies smaller than 0.7 keV. Thus, if this dense clumps at subvirial temperatures are very cold ($\lesssim 1$ keV), high-z clusters might be less biased by the presence of a gas at low temperatures, but at the same time it will be more difficult to detect these substructures given the cosmological dimming. Shallow observations are also more prone to be biased by undetected clumps. In Figure \ref{bias} we present the bias on the gas density due to undetected substructures at subvirial temperatures. 

\begin{figure*}
\begin{center}
\hbox{
\psfig{figure=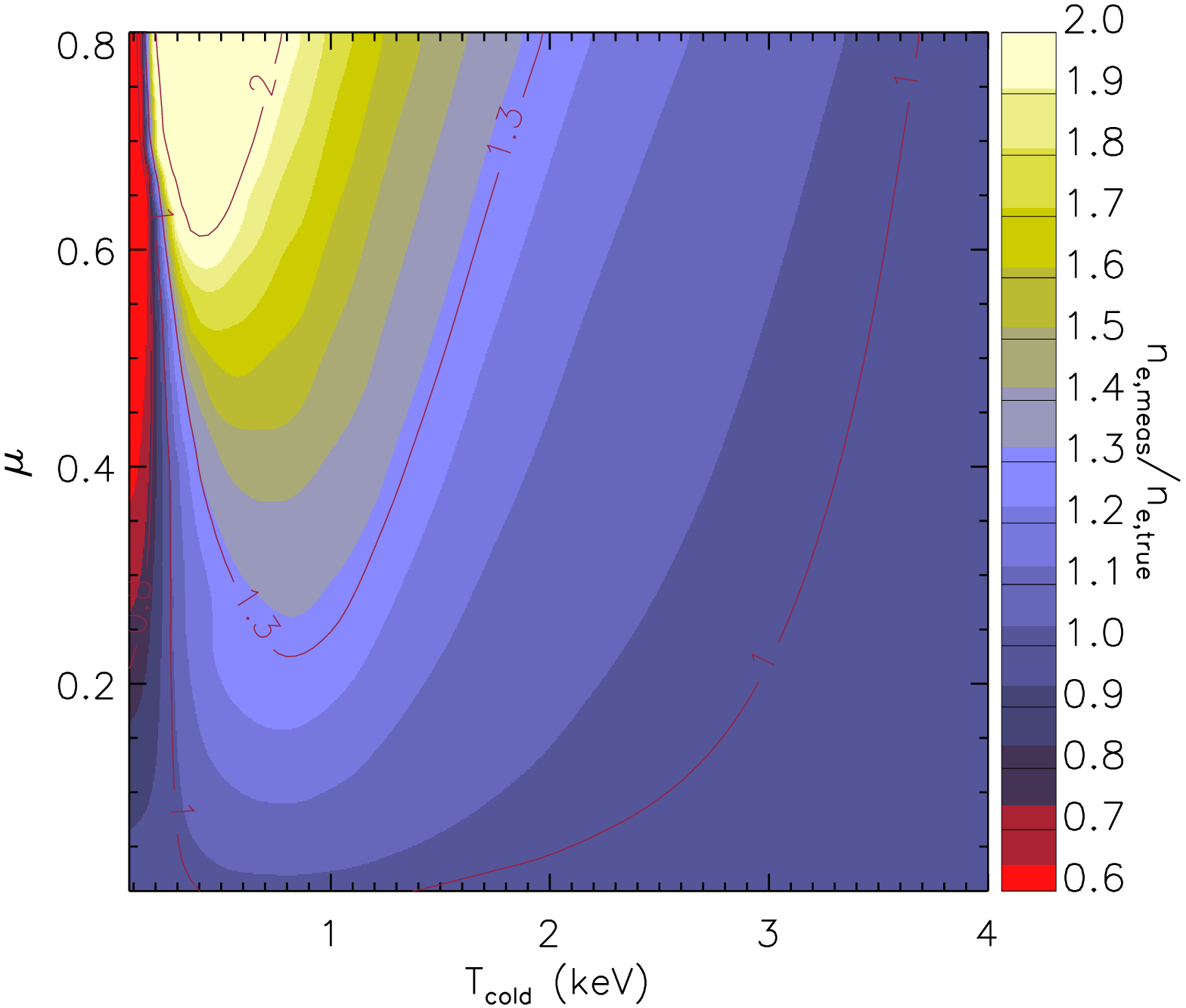,width=0.5\textwidth}
\psfig{figure=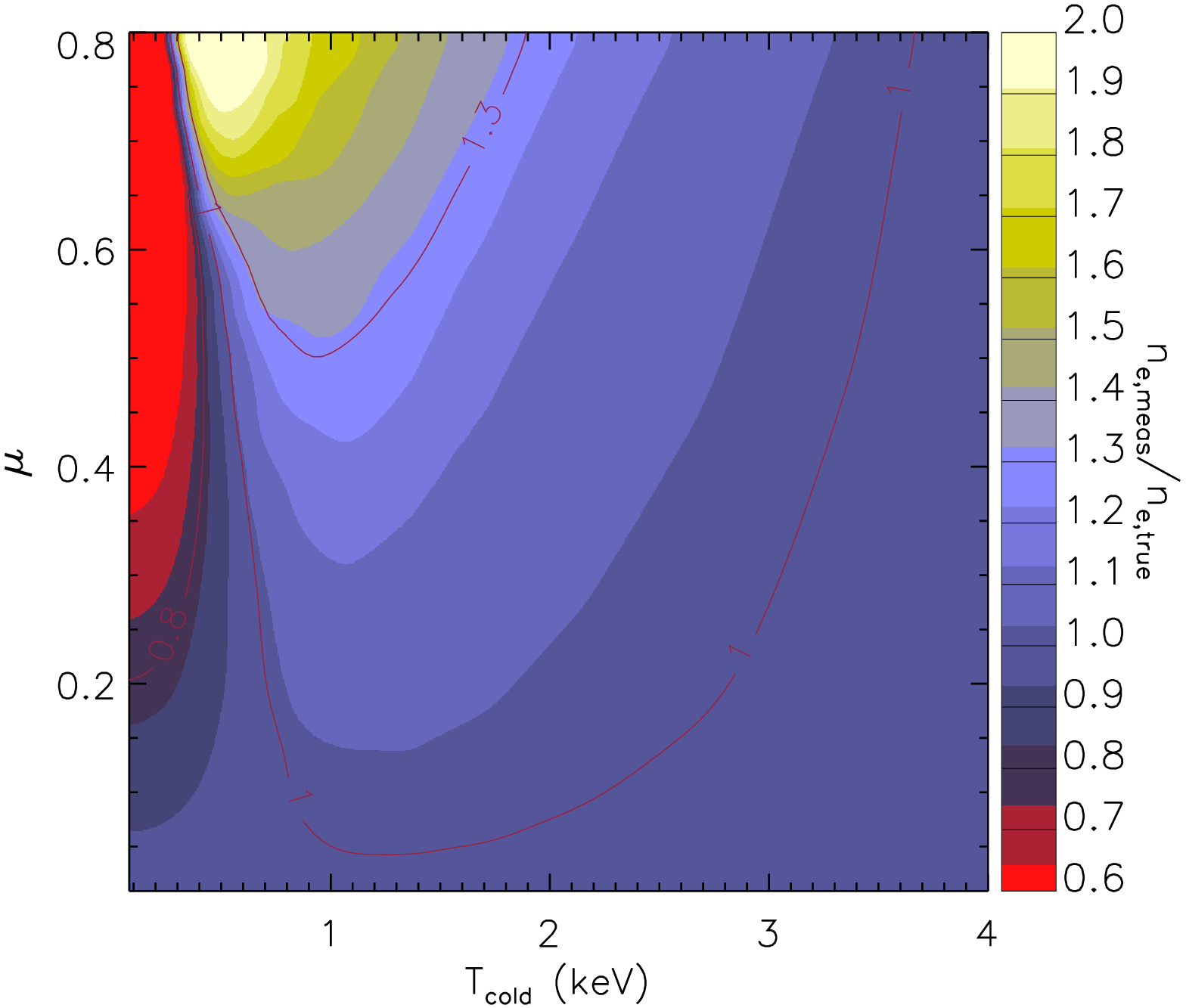,width=0.5\textwidth}
}
\caption[]{Bias on the gas density due to a multi-temperature gas distribution of the ICM. We fitted an absorbed thermal emission model at a single emission temperature $T_{\rm spec}$ on mock spectra of a two-phase plasma at temperatures $T_{\rm hot}$ and $T_{\rm cold}$ and in pressure equilibrium, with gas mass fraction of the cold gas component $\mu$. We assume that the gas mass-weighted temperature of this two-phase plasma (as measured e.g. by SZ) is constant, i.e. $T_{\rm mw}=4$ keV. The left (right) panel refers to an observation redshift of 0.05 (1).}
\label{evderydddg}
\end{center}
\end{figure*}                           

The qualitative picture emerging by this toy model is that the presence of undetected substructures might bias upwards the measured gas density, with a magnitude larger for shallow observations and low redshifts. Nevertheless, by dividing our sample into two subsamples, clusters with total net counts smaller and larger than the median value of the whole sample, respectively, and repeating our stacking procedure, we do not find any significant difference in the recovered physical parameters, which might arise given the different impact of substructures upon the observation time. While the superb angular resolution of {\em Chandra} is essential for detecting small-scale clumps and distinguishing the X-ray emissions arising from clumps and diffuse components, we caution the reader that it possible we might actually missing some substructures even in deep observations, given their low signal-to-noise in the outskirts, which could enhance the X-ray emission and hence the gas density. 

Moreover, inhomogeneities in the gas distribution can also lead to the overestimate of the observed gas density \citep {nagai2011,zhuravleva2013,roncarelli2013}, which in turn introduces biases in global cluster properties, such as the gas mass fraction \citep{battaglia2012} and the low-scatter X-ray mass proxy, such as $Y_X \equiv M_{\rm gas} T_X$ \citep{khedekar2013}. At $R_{200}$ hydrodynamical simulations predict a gas clumping factor in the range $\sim 1.3-2$ \citep{nagai2011,zhuravleva2013,battaglia2012}. Gas clumping factor inferred from these simulations is in reasonable agreement with observations \citep{morandi2013b,morandi2014}. If they are not properly understood and modeled, these non-equilibrium processes can limit the use of galaxy clusters as cosmological probes.

Finally, in order to understand how a multiphase structure of the ICM can bias X-ray observations, we present a simple toy problem and we analyze the effect of multi-temperature distribution on the physical observables (i.e. X-ray spectroscopic temperature and gas density). A cold phase of the ICM would systematically bias down X-ray spectroscopic temperatures with respect to the average (gas mass-weighted) temperature e.g. probed through SZ. We consider a two-phase gas at temperatures $T_{\rm cold}$ and $T_{\rm hot}$, with gas masses $m_{\rm cold}$ and $m_{\rm hot}$, respectively, and in pressure equilibrium. Hence the gas mass-weighted temperature $T_{\rm mw}$ reads:
\begin{equation}\label{mix1}
T_{\rm mw}=(1-\mu)T_{\rm hot}+\mu T_{\rm cold}\ , \quad  \mu=m_{\rm cold}/(m_{\rm cold}+m_{\rm hot})
\end{equation}
with $T_{\rm mw}=T_{\rm mw}(T_{\rm hot},T_{\rm cold},\mu)$, $T_{\rm cold}\lesssim T_{\rm mw}\lesssim T_{\rm hot}$. 

We generated mock spectra of the previous plasma by means of absorbed APEC emission models at temperatures $T_{\rm hot}$ and $T_{\rm cold}$, and under the boundary condition that gas mass-weighted temperature (Equation \ref{mix1}) is constant, i.e. $T_{\rm mw}=4$ keV. The hydrogen column density $N_H$ is fixed to $10^{20}\; \mbox{cm}^{-2}$. The current emission model is then folded through response curves (ARF and RMF) of {\em Chandra}. 

Since $T_{\rm mw}$ is constant, the problem to recover the spectral temperature $T_{\rm spec}$ of the two-phase plasma is fully described via two independent variables, e.g. $T_{\rm cold}$ and $\mu$, while $T_{\rm hot}$ is a dependent variable which can be inferred via Equation \ref{mix1}. Thus, $T_{\rm spec}=T_{\rm spec}(T_{\rm cold},\mu)$, where we measured the spectral temperature by fitting an absorbed APEC emission models at a single-temperature $T_{\rm spec}$ on the mock spectra of the two-phase plasma. The final goal is to infer the bias on the gas density, which arises since the spectroscopic temperature $T_{\rm spec}$ is then used to convert observed X-ray counts into an emission measure (and thus gas density) via a single-temperature absorbed thermal model. In Figure~\ref{evderydddg} we present the results of the bias on the gas density for a grid of values of $T_{\rm cold}$ and $\mu$.

We can observe that the gas density on the two-phase plasma is biased with respect to the true density. Since the primary emission mechanism in X-ray clusters is thermal bremsstrahlung and emission lines, the spectral fit is indeed guided, primarily, by the exponential bremsstrahlung cutoff at high energies, the iron line at energies $\sim7/(1+z)$ keV, and the lower energy ($\lesssim 2$ keV) emission lines. Complications arise since the true spectrum of the astrophysical source must be convolved with the instrumental response of the X-ray telescope. {\em Chandra} effective area, for example, is small for photon energies smaller than 0.6 keV or greater than 7 keV. Hence, the presence of a cool phase at temperature $0.5\lesssim T \lesssim 2$ keV in the ICM will typically bias down the spectroscopic temperature (single-temperature absorbed thermal model), since the instrument is very sensitive to the low energy features, such as low energy emission lines. In Figure~\ref{evderydddg} we can see that the gas density is biased upwards as long as $\sim T_{\rm cold}/(1+z)\gesssim 0.25$ keV, a with non-negligible dependency on the observation redshift, $\mu$ and $T_{\rm cold}$. Counterintuitively, for low temperatures $\sim T_{\rm cold}/(1+z)\lesssim 0.25$ keV and larger values of the cold phase fraction $\mu \gesssim 0.2$, the measured gas density is biased downwards of $\sim0-20\%$. This is due to the fact that most of the emission captured by the X-ray telescope arises from the hot phase of the ICM, with temperature $T_{\rm hot}\gesssim T_{\rm mw}$ for $T_{\rm cold}/(1+z)\lesssim 1$ keV. On the contrary, most the emission of the cold phase would be at very low energies, where the {\em Chandra} effective area is small, and related to emission lines rather than bremsstrahlung. This bias is increasing for higher observation redshifts.

A multiphase structure of the ICM would retain signatures of the 'melting pot' in the virialization region, where infalling clumps of matter and gas are predicted to have, typically, higher density and cooler temperature than their surroundings. Current measurements via X-ray telescopes, which customarily hinge on a single-temperature absorbed thermal model, can be thus biased. Future instruments (e.g., the upcoming Astro-H and an envisaged Athena-like mission) will have enough collecting area and spectral resolution to better distinguish multi-temperature structures spectrally. Moreover, more detailed simulations (e.g. hydrodynamical simulations) including these effects might be needed to address the impact of these biases.

\subsection{Other sources of systematics}\label{sys453hrhrh34}

We checked the effect of the presence of a temperature profile on the measured emission measures. The conversion from ACIS-I count rate to emission measure has been determined via an absorbed thermal emission model folded with the AICS-I response. This conversion factor was inferred for each cluster by using the global spectroscopic temperature in the range $0.15-0.75\, R_{200}$ rather than the temperature profile. As we previously (\S\ref{laoa3}) pointed out, the conversion from ACIS-I count rate to emission measure is highly insensitive to the temperature (for $T> 2$ keV). The conversion factor changes at most by 8\% for low-temperature ($T\sim 3$ keV) clusters, which translates into a bias $\lesssim 4\%$ on the gas density, while it is negligible ($\lesssim 1\%$) for hot ($T\gesssim 6$ keV) systems. To further check the effect of the presence of a temperature profile on the measured stacked emissivity and hence gas density, we generated mock event files for each cluster under these assumptions: i) the gas density distribution for each cluster is self-similar and fixed to that measured via our stacking method as described in \S\ref{laoa3}; ii) we assume the expression for the self-similar temperature profile that \cite{vikhlinin2006} showed to reproduce well the temperature gradients in nearby relaxed systems, with normalization fixed to the spectral temperature of each source. We repeated the stacking procedure described in \S\ref{laoa3} by assuming first isothermality and then by accounting for the temperature profile. We verified that the recovered physical parameters are biased downwards in a negligibly way ($\lesssim 2\%$ for the emission measure) by our method. 

As a further test, we compared our gas mass at $R_{500}$ (recovered via deprojection of individual emission measure profiles) with the results of \cite{maughan2008}, whose sample shares 95 objects in common with ours. We also compared with \cite{martino2014}, whose sample contains 32 objects in common with us. We found no significant differences between the recovered gas masses (Appendix \ref{maugh1}). We point out that these individual gas masses profiles have been recovered only for the purpose of this comparison, since we actually use stacked profiles across the paper.

\begin{figure*}
\begin{center}
\hbox{
\psfig{figure=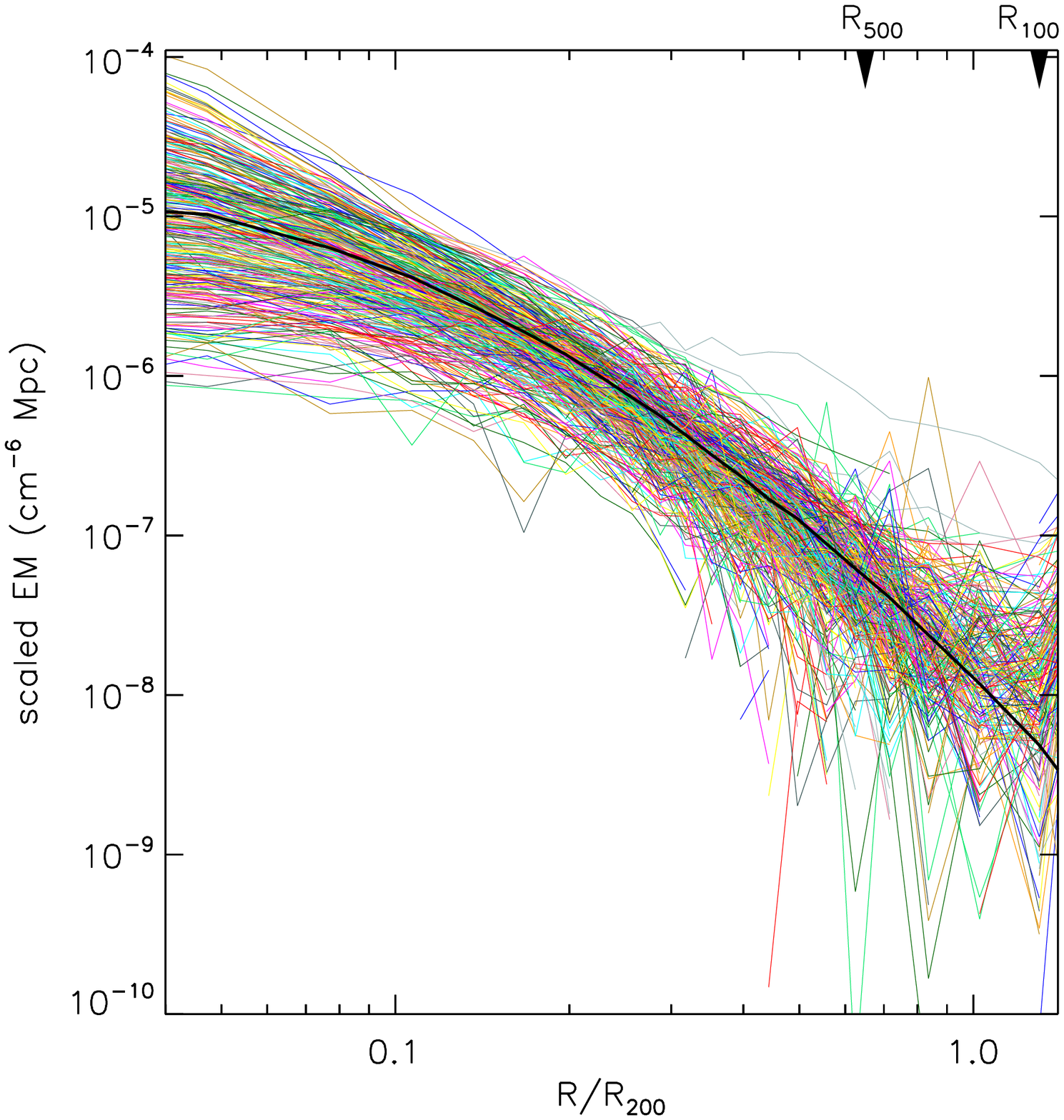,width=0.5\textwidth}
\psfig{figure=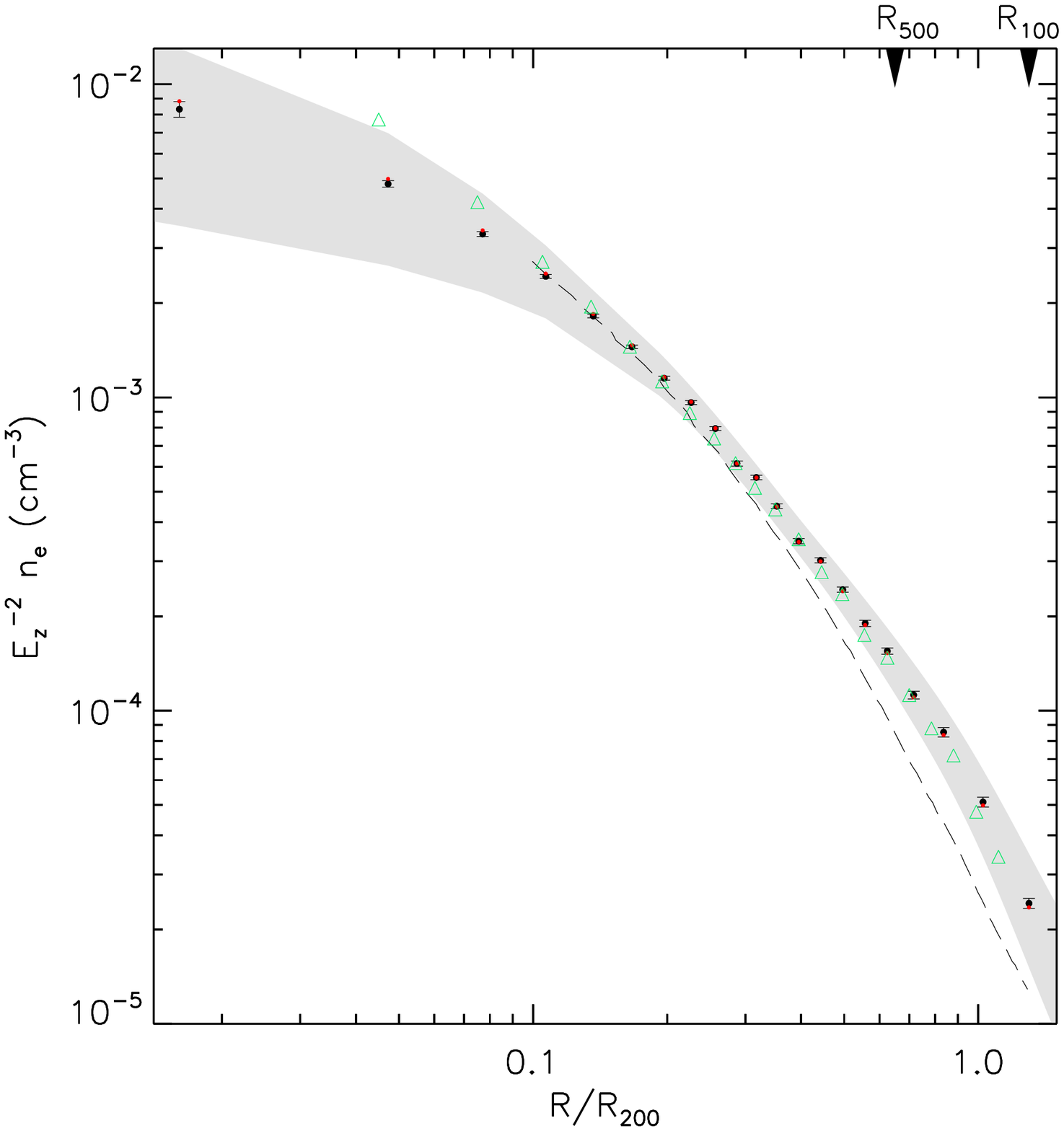,width=0.5\textwidth}
}
\caption{Left: The {\em Chandra} emission measure profiles of all clusters in the sample, re-scaled according the self-similar model (\S\ref{selfsim}). A straight line is drawn between the data points for each cluster for a better visualization of the profiles. The thick black solid line represents the median emission measure profile. Right: average gas density profile for the whole sample. The points with errorbars represent the median values, while the red points represent the weighted average profile. Note that the errorbars refer to the errors on the median emission measure profiles, while the gray shaded region refers to the intrinsic scatter (68.3\% confidence level), which amounts to $\sim20\%$ ($\sim30\%$) at $R_{500}$ ($R_{500}$). The long dashed line represents the predictions from the hydrodynamic numerical simulations of \citet{roncarelli2006}, while the green triangles represents the universal gas density profile from \citet{eckert2012} based on {\em ROSAT} data.}
\label{evderyddd}
\end{center}
\end{figure*}

Next, we compared the stacked emission measure profile by using median and weighted average of the emission measure profiles, and we did not find any statistically significant difference for $R\gesssim 0.05R_{200}$. Similar conclusions hold for the gas density (Figure \ref{evderyddd}). This is a consequence of the fact that the errors (measurement and systematic) are symmetric. However, a weighted-average would significantly underestimate the final uncertainties on the stacked emission measure, since it accounts only for the measurement errors but it neglects the intrinsic scatter of the emission measure profiles.  Hereafter, we will refer to the results recovered via the median of the emission measure profiles.

Finally, we estimated the bias due to uncertainty in the $M_{500}-Y_{X,500}$ relation \citep{sun2009} on the desired physical parameters due to the presence of non-thermal pressure. The latter can indeed bias downwards the hydrostatic masses through which the $M_{500}-Y_{X,500}$ relation has been calibrated. We define the mass bias $b$ between the true and hydrostatic mass $M_{hydro}$, with both masses defined at a fixed density contrast of 500. Hence we have that $M_{true}=M_{hydro}/(1-b)$. We further assume an universal gas density profile as determined in our work (\S\ref{sys453b}), and the expression for the self-similar temperature profile from \cite{vikhlinin2006}. Thus, by changing the bias parameter $b$ and correcting accordingly the normalization of the $M_{500}-Y_{X,500}$ relation, we can infer the $R_{500}=R_{500}(T,b)$ relation from Equation \ref{e.ym}, which can adequately described by the following expression:
\begin{equation}\label{e.ym2}
 R_{500}  = E_{z}^{-1} 1095\; {\left(\frac{T_X}{\rm 6\, keV}\right)}^{0.44} \,10^{0.37\,b}\ {\rm kpc}
\end{equation}
The bias on the true gas density becomes $n_{e,true}/n_{e,meas}\simeq {(R_{500}(b)/R_{500,meas} )}^{-\alpha}$. Note that the bias parameter $b$ refers to the non-thermal pressure support at $R_{500}$, since the $M_{500}-Y_{X,500}$ relation has been calibrated at this overdensities. For larger radii we do not expect a larger impact of the non-thermal pressure, since we used the $Y_{X,500}$ estimator in order to infer $R_{500}$; hence $R_{200}$ and $R_{100}$ have been statistically determined by assuming an NFW distribution for the total matter (\S\ref{selfsim}). From Equation \ref{e.ym2} and for $b\sim 0.1$ \citep[see, e.g.][]{mahdavi2008,morandi2012a,morandi2012b} the presence of non-thermal pressure can bias upwards the gas density of $\sim15(25)\%$ at $R_{500}(R_{200})$.

\section{The outskirts of cluster: gas density}\label{sys453b}
Results for the gas density, gas density slope $\beta=-1/3\,d\log(n_e)/d\log(r)$ and gas fraction $f_{\rm gas}(<R) =M_{\rm gas}(<R)/M_{\rm tot}(<R)$ for our sample are presented in Table~\ref{tab:1eggw}. We divided the data into subsamples to investigate the dependency on the redshift $z$, cluster temperature $T$, cluster morphological type (relaxed/unrelaxed). We conventionally defined relaxed (unrelaxed) system with $w<10^{-2}$ ($w>10^{-2}$). We remember that the centroid shifts and central density are anti-correlated (Figure \ref{w-cc}), being both a proxy of the presence of a cool-core: thus the chosen threshold of $w$ roughly demarcates the division between CC/NCC. 

The quoted uncertainties reflect the total error budget (measurement error + intrinsic scatter) by means of MC randomizations.

In order to gauge the bias implicit in our spherical modeling, we further repeated the X-ray analysis by restricting the stacking procedure (\S\ref{laoa3}) in two sectors ($f_1$ and $f_2$ in Table~\ref{tab:1eggw}). The sector $f_2$ has position angle fixed to the major axis of the X-ray brightness distribution and with an aperture of 60 degrees, i.e. encompassing the cosmic filament (see \S\ref{morph} for further discussion), while the other ($f_1$) characterizes the remaining volumes of the cluster.

\begin{table*}
\scriptsize
\begin{center}
\caption{Gas density, gas density slope $\beta=-1/3\,d\log(n_e)/d\log(r)$ and gas fraction at different overdensities. The quoted errors refer to the errors on the median emission measure profiles.}
\begin{tabular}{c@{\hspace{.8em}} c@{\hspace{.8em}} c@{\hspace{.8em}} c@{\hspace{.8em}}c@{\hspace{.8em}} c@{\hspace{.8em}} c@{\hspace{.8em}} c@{\hspace{.8em}} c@{\hspace{.8em}}   c@{\hspace{.8em}} }
\hline \\
&  & $\Delta=500$ & & & $\Delta=200$ &  & & $\Delta=100$ & \\
\hline \\
sample & $E_z^{-2} n_e$ & $\beta$ & $f_{\rm gas}$  &$E_z^{-2} n_e$ & $\beta$ & $f_{\rm gas}$  &$E_z^{-2} n_e$ & $\beta$ & $f_{\rm gas}$ \\
&$(10^{-4}\mbox{cm}^{-3})$ & &  & $(10^{-5}\mbox{cm}^{-3})$ & & & $(10^{-5}\mbox{cm}^{-3})$ & & \\
\hline \\
$all$ &  $  1.304\pm  0.034$ & $  0.677\pm  0.018$ & $  0.130\pm  0.002$ & $  4.780\pm  0.199$ & $  0.940\pm  0.018$ & $  0.150\pm  0.004$ & $  2.161\pm  0.130$ & $  1.280\pm  0.060$ & $  0.158\pm  0.005$  \\
$T_1$ &  $  1.310\pm  0.038$ & $  0.683\pm  0.017$ & $  0.131\pm  0.002$ & $  4.744\pm  0.208$ & $  0.936\pm  0.018$ & $  0.151\pm  0.006$ & $  2.116\pm  0.134$ & $  1.326\pm  0.060$ & $  0.159\pm  0.006$  \\
$T_2$ &  $  1.275\pm  0.068$ & $  0.646\pm  0.026$ & $  0.127\pm  0.003$ & $  5.012\pm  0.373$ & $  0.945\pm  0.028$ & $  0.148\pm  0.008$ & $  1.827\pm  0.160$ & $  1.375\pm  0.077$ & $  0.157\pm  0.008$  \\
$T_3$ &  $  1.299\pm  0.095$ & $  0.641\pm  0.031$ & $  0.128\pm  0.003$ & $  4.933\pm  0.473$ & $  0.898\pm  0.031$ & $  0.150\pm  0.011$ & $  2.247\pm  0.186$ & $  1.130\pm  0.072$ & $  0.161\pm  0.011$  \\
$z_1$ &  $  1.373\pm  0.047$ & $  0.642\pm  0.051$ & $  0.130\pm  0.002$ & $  4.781\pm  0.318$ & $  1.004\pm  0.064$ & $  0.154\pm  0.005$ & $  2.308\pm  0.251$ & $  1.121\pm  0.141$ & $  0.163\pm  0.008$  \\
$z_2$ &  $  1.289\pm  0.074$ & $  0.681\pm  0.051$ & $  0.133\pm  0.003$ & $  5.233\pm  0.344$ & $  1.079\pm  0.127$ & $  0.156\pm  0.009$ & $  2.281\pm  0.214$ & $  1.299\pm  0.121$ & $  0.161\pm  0.007$  \\
$z_3$ &  $  1.287\pm  0.081$ & $  0.609\pm  0.029$ & $  0.125\pm  0.003$ & $  4.639\pm  0.407$ & $  0.943\pm  0.036$ & $  0.144\pm  0.009$ & $  2.144\pm  0.192$ & $  1.238\pm  0.069$ & $  0.155\pm  0.009$  \\
$w_1$ &  $  1.260\pm  0.051$ & $  0.657\pm  0.046$ & $  0.129\pm  0.002$ & $  4.937\pm  0.281$ & $  0.922\pm  0.081$ & $  0.148\pm  0.006$ & $  2.213\pm  0.206$ & $  1.212\pm  0.174$ & $  0.156\pm  0.006$  \\
$w_2$ &  $  1.324\pm  0.050$ & $  0.676\pm  0.019$ & $  0.130\pm  0.002$ & $  4.805\pm  0.260$ & $  0.941\pm  0.018$ & $  0.152\pm  0.006$ & $  2.261\pm  0.145$ & $  1.273\pm  0.058$ & $  0.161\pm  0.007$  \\
$f_1$ &  $  1.215\pm  0.038$ & $  0.692\pm  0.022$ & $  0.125\pm  0.002$ & $  4.289\pm  0.246$ & $  1.056\pm  0.029$ & $  0.141\pm  0.005$ & $  1.958\pm  0.162$ & $  1.369\pm  0.066$ & $  0.148\pm  0.006$  \\
$f_2$ &  $  1.401\pm  0.046$ & $  0.625\pm  0.046$ & $  0.135\pm  0.002$ & $  5.051\pm  0.251$ & $  0.853\pm  0.033$ & $  0.162\pm  0.004$ & $  2.690\pm  0.133$ & $  0.985\pm  0.077$ & $  0.168\pm  0.006$  \\
\hline \\
\end{tabular}
\begin{flushleft}
Notes on the sample symbols: $all$: whole sample; $T_1:kT<6$ keV; $T_2:6<kT<8$ keV;  $kT>8$ keV; $z_1:z<0.3$; $z_2: 0.3<z<0.6$; $z_3: z>0.6$; $w_1: w<10^{-2}$;  $w_2: w>10^{-2}$; $f_1:$ excluding the filament; $f_2:$ including only the filament.\\
While the quoted errors refer to the errors on the median, we also report the value of the intrinsic scatter for the whole sample at $R_{500}$($R_{200},R_{100}$): for $n_e$ we have $\sim 20\%(30\%,38\%)$; for $\beta$ we have $\sim 8\%(12\%,15\%)$; for $f_{\rm gas}$ we have $\sim 15\%(25\%,28\%)$.
\end{flushleft}
\label{tab:1eggw}
\end{center}
\end{table*}

\subsection{Gas density slope and normalization}
Our analysis on these clusters (Figure \ref{evderyddd}) reveals that the self-similar model is a tenable across a wide redshift and mass range. Different from some recent {\em Suzaku} results, and confirming previous evidence from {\em ROSAT} and {\em Chandra} \citep{vikhlinin1999,vikhlinin2006,eckert2012}, our average emission measure profile translates into a steep density profiles. We observe a general trend of steepening in the radial profile of the gas density beyond $0.3\, R_{200}$, with $\beta \sim 0.940\pm  0.018$ at $R_{200}$. This steepening is in agreement with previous works by {\em ROSAT} \citep{vikhlinin1999}, but at odds with some recent {\em Suzaku} X-ray observations, where the electron density decreases steadily with radius, approximately following a power-law model with $\beta\sim0.7$ \citep{simionescu2011,kawaharada2010}. 

\begin{figure*}
\begin{center}
\hbox{
\psfig{figure=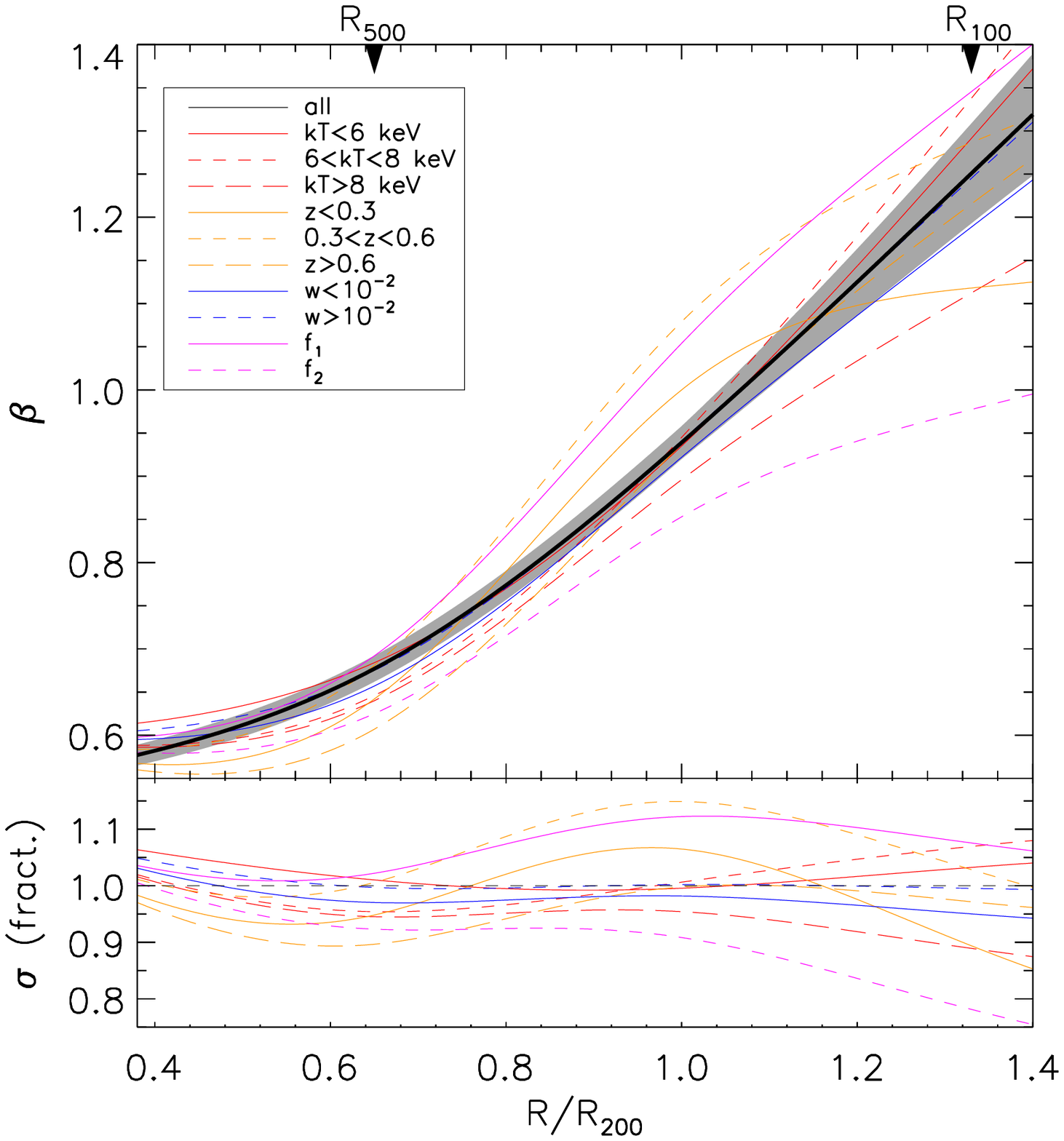,width=0.5\textwidth}
\psfig{figure=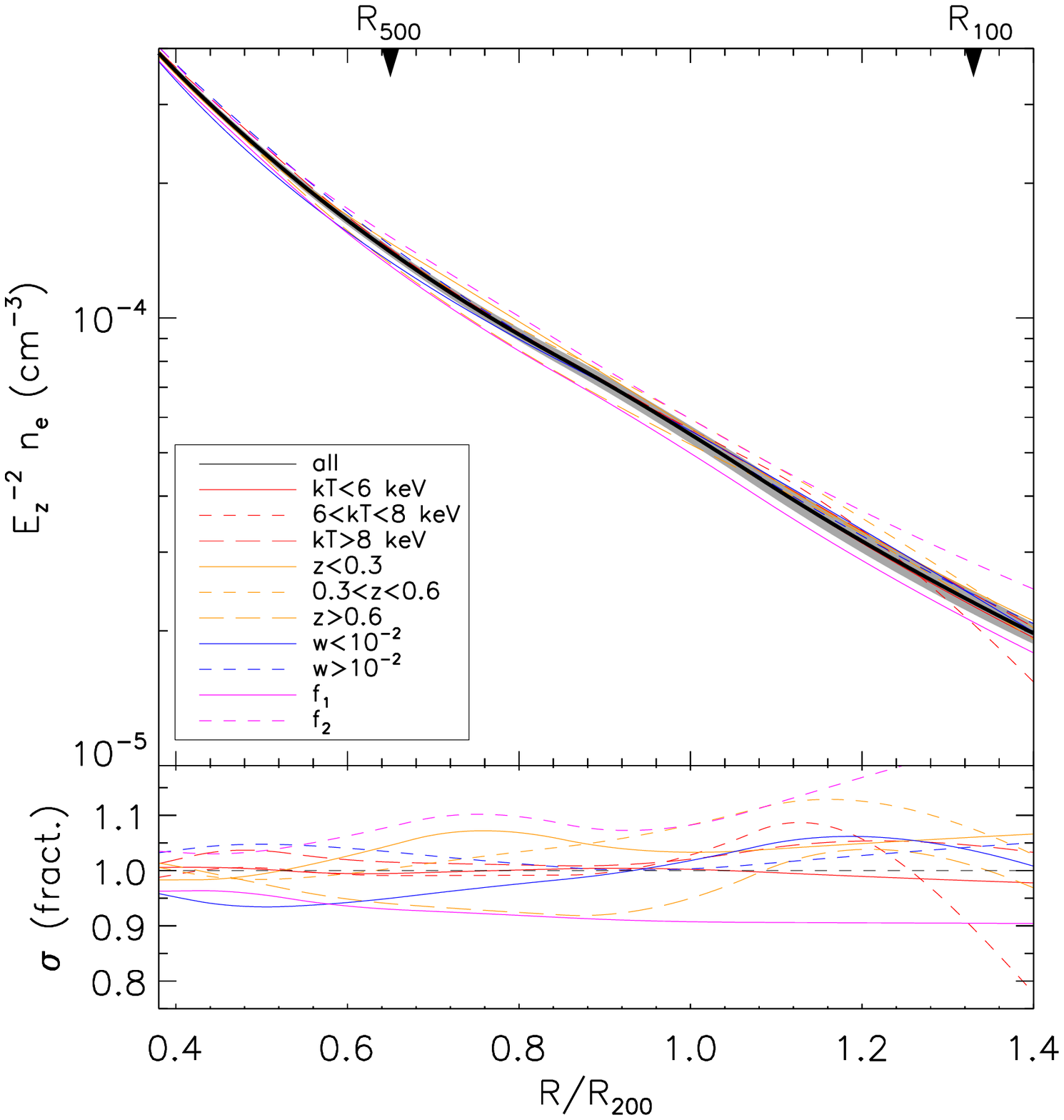,width=0.5\textwidth}
}
\caption{Slope of the gas density $\beta=-1/3\,d\log(n_e)/d\log(r)$ (left panel) and electron density $n_e$ as a function of the radius for different populations of the cluster sample. The solid/dashed lines represent the median values for the different subsamples, while the 1-$\sigma$ errors for the total sample are represented by the gray shaded region. The panels at the bottom represent the fractional variation of the profiles of the subsamples  with respect to the whole sample. In order to improve the readability of the figures, we interpolated the data via a least-squares constrained spline approximation so as to have continuous functions. }
\label{bias335}
\end{center}
\end{figure*}

By using the centroid shift as a proxy of the dynamical state of clusters and the presence of a cool-core (Figure \ref{w-cc}), we did not detect any systematic difference between cool-core/relaxed ($w<10^{-2}$) and non-cool core/unrelaxed ($w>10^{-2}$) clusters at $R_{500}$ (Figure \ref{bias335} on the right). We found that NCC clusters do not have on average a higher density than CC systems, which might arise due to larger clumping factor and/or merging events in disturbed objects. In this respect \cite{eckert2012} noted a clear distinction between the two classes in cluster outskirts across all radii, out to $R_{200}$, in moderate disagreement with the present analysis. As pointed out by \cite{eckert2012}, this difference in their analysis could be explained by an inaccurate determination of $R_{200}$ for unrelaxed clusters in the  sample. Indeed, they use the $R_{200}-T$ scaling relation computed under the assumption of hydrostatic equilibrium. On the contrary, in order to compute $R_{200}$, in the present analysis we used the $Y_{X,500}-M_{500}$ relation, which should be nearly insensitive to differences between relaxed and unrelaxed objects \citep{kravtsov2006}.

We also do not detect evolution of the normalization of the gas density; however, we find hints of evolution of the slope of the recovered density profile slope at $R_{200}$ is $\beta=1.004\pm  0.064$ ($1.079\pm  0.127$, $0.943\pm  0.036$) for $z<0.3$ ($0.3<z<0.6$, $z>0.6$, Figure \ref{bias335} on the left). This difference can be the result of the evolution of energy feedback processes from e.g. AGNs, supernovae and star formation as a function of the redshift. Note that current {\em Suzaku} and {\em ROSAT} constraints are limited to very low redshift, due to their poor spatial resolution. In this respect, \cite{eckert2012} performed a stacking of the density profiles of a sample of $\sim 30$ clusters at low redshift observed through {\em ROSAT}, finding $\beta=0.890 \pm 0.026$ at $R_{200}$, in agreement within $1.7\sigma$ with our results on the low-redshift subsample.

A posteriori, our findings on different subsamples suggest that our assumption of self-similarity holds for the ICM in the virialization region in clusters, regardless their dynamical states, temperature and redshift.

Finally, at $R_{500}$ the intrinsic scatter of the density profiles is $\sim20\%$, in agreement with previous studies \citep{vikhlinin2006,eckert2012}. At $R_{200}$, the electronic density inferred on the whole sample is $E_z^{-2} n_e=(4.780\pm  0.199)\times 10^{-5}\mbox{cm}^{-3}$, with an intrinsic scatter of $\sim$30\%: this value is in agreement with previous {\em ROSAT} measurements, i.e. $E_z^{-2} n_e=(4.6\pm  0.5)\times 10^{-5}\mbox{cm}^{-3}$, with a 25\% scatter \citep[][]{eckert2012}.

\subsection{Comparison with simulations}\label{compsim}
Comparing our {\em Chandra} density profile with numerical simulations including cooling, star formation and supernovae feedback \citep{roncarelli2006}, we found that the latter fail to reproduce the observed shape of the density profile, predicting density profiles that are significantly too steep and with lower normalization compared to the data (Figure \ref{evderyddd}). This might indicate that non-gravitational effects are important well outside the core region, whereas in the above simulation Roncarelli and collaborators did not include AGN feedback and/or preheating. Indeed, recent works \citep[][]{mathews2011} indicate that feedback mechanisms may be responsible for the deficit of baryons in cluster cores, smoothing the accretion pattern and leading to a flatter gas distribution. This is in agreement with the findings of \cite{morandi2014,eckert2013a}, who also argued for an entropy excess significantly beyond $R_{500}$, although we caution the reader that we do not have spectroscopic radial temperature/entropy profiles to independently validate this scenario. 

The normalization of simulations is also too small: it is indeed well known that these simulations overpredict the stellar fraction in clusters \citep{kravtsov2012}. This large reservoir of cold baryons in simulations might potentially explain the discrepancy in the X-ray gas density normalization between the present observations and simulations. Consequently, a detailed treatment of gas cooling, star formation, AGN feedback, and consideration of gas clumping is required to have realistic models of the outer regions of clusters.

Alternatively, it is possible that X-ray are strongly biased by the presence of a cold phase and/or clumpy gas distribution, or undetected substructures, enhancing the X-ray emission and hence the gas density. The CDM scenario indeed predicts a picture where clumps at subvirial temperature are infalling along filaments and accreting onto the cluster outskirts. Given the low signal-to-noise and/or poor spatial resolution of the X-ray/SZ telescope, it might be impossible to mask out these substructures e.g. in the X-ray analysis, and they might remain undetected. While Roncarelli and collaborators eliminated in the simulations the densest clumps (the one per cent of the volume that corresponds to the densest SPH particles) that dominate the density and surface brightness in the outskirts, this empirical method might not fully capture the procedure of masking bright isolated regions from the analysis of observed clusters. This procedure indeed depends on the particulars of the observations, e.g. integration time, X-ray luminosity of the cluster and clumps versus the CXB, redshift of the observation, the value of the Galactic neutral hydrogen absorption, spatial resolution of the X-ray telescope (see \S\ref{bias11} and Figure \ref{bias} for further discussion).

We then compared our observational results with the properties of simulated galaxy clusters produced with the adaptive mesh refinement code \texttt{ENZO} \citep{vazza2013}. We report that the gas density profiles of simulations is roughly in agreement with observations for $R \gesssim R_{\rm 500}$, the former predicting a slope of $\sim0.97$ at the virial radius \citep[see][]{eckert2012}. A reasonable agreement with theoretical expectations is found when we compare our findings with the hydrodynamical numerical simulations of galaxy cluster formation that include radiative cooling, star formation, metal enrichment and stellar feedback from \cite{nagai2007a}. These simulations indicate gas density profiles with $\beta\sim 0.9$ at $R_{200}$.

\subsection{Unveiling the ongoing large-scale structure formation scenario}\label{lsst3}
Since baryons in the outskirts bear the signature of the continuous three-dimensional accretion from surrounding filaments, measuring the magnitude of non-equilibrium processes such as clumpy gas distribution, the presence of substructures, asphericity in the gas distribution and complex accretion patterns provides a neat way to assess the virialization degree achieved by the ICM.

In particular, numerical simulations indicate that the infall of material into the most massive dark matter haloes is not spherical but is expected to be preferentially funneled through the cosmic filaments where the clusters are embedded. The cluster mass haloes would indeed acquire most of their mass from major mergers along the filaments, hence leading to an alignment between the major axis of the host halo and the large-scale filament \citep{brunino2007}. Therefore an ellipsoidal gas and DM distribution appears to be a direct outgrowth the ongoing accretion scenario \citep{limousin2013}, retaining the fingerprints of the topology of the cosmic structures. Independent X-ray observations confirmed that there is a substantial alignment between the major axis of the brightness distribution and the large-scale filament \citep{morandi2014}. Motivated by this cosmic structure formation scenario, we use the position angle via the moments of the surface brightness distribution (\S\ref{morph}) as a proxy of the direction of the large-scale filament where the cluster is embedded. 

In order to gauge the bias implicit in our spherical modeling, we repeat the X-ray stacking analysis extracting surface brightness profiles for the $0.7-2$ keV band for two sectors excluding and encompassing the cosmic filament ($f_1$ and $f_2$, respectively). With respect to the azimuthally-averaged density profile, the sector $f_1$ ($f_2$) shows a density lower (higher), the net effect being to systematically shifting downwards (upwards) the gas density profile, with a slope which is flatter when the filament is included (Figure \ref{bias335}). The difference in gas density normalization between the two sectors is $\sim30\%$ at $R_{100}$. Thus, beyond $R_{500}$, galaxy clusters deviate significantly from spherical symmetry, with only small differences between relaxed and disturbed systems, with the gas located mostly along preferential directions (i.e., filaments).

In light of our hypothesis that the elongation of the ICM tracks the large-scale filament, the gas density along the directions of filaments where the cluster accretes clumpy and diffuse materials would then be enhanced in qualitative agreement with our analysis. The bias on the X-ray emissivity due to subclumps falling into the DM potential well under the pull of gravity should be also more pronounced along the direction of the cosmic filaments, if these substructures remain undetected in the X-ray analysis. In this respect, as previously pointed out also by \cite{eckert2012}, we argued that the shallow density profiles observed in some clusters by {\em Suzaku} might be induced by observations in preferential directions (e.g. along filaments), since some these observations have been carried along narrow arms, and do not reflect the typical behavior of cluster outer regions. 

Finally, we point out that that in Figure \ref{bias335} the differences between the $f_1$ and $f_2$ sectors originate also within $R_{500}$, in regions where you should naively expect a negligible impact due to the cosmic filaments. Remembering that we recovered the emission measure profiles in circular annuli, this trend in these inner volumes is probably due to (mostly) triaxial gas distribution and not to the real presence of filaments. It is indeed well known that the ICM distribution follows isopotential surfaces approximated by concentric ellipsoids of decreasing axial ratio towards the outer volumes, the dynamics of the gas being driven by the gravitational potential well of the DM \citep{morandi2010a,morandi2012a,morandi2012b}. However, in the virialization region ($R\gesssim R_{500}$) the isopotential surfaces of the ICM are expected be nearly spherical, and any differences between the $f_1$ and $f_2$ sectors must genuinely originate from the impact of the cosmic filaments rather from triaxiality.

\section{Gas fraction}\label{syeebse2}
We studied the hot gas mass fraction $f_{\rm gas}(<R) =M_{\rm gas}(<R)/M_{\rm tot}(<R)$ via stacking of our archival {\em Chandra} data. We present X-ray measurements of the hot gas fraction out to $R_{100}$ by addressing the impact of dynamical state, temperature cluster, evolution with $z$ and asphericity.

\begin{figure}
\begin{center}
\hbox{
\psfig{figure=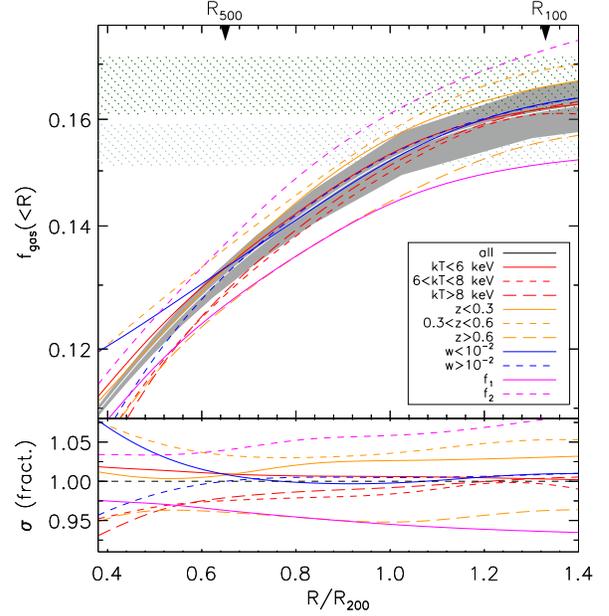,width=0.5\textwidth}
}
\caption{Cumulative gas fraction $f_{\rm gas}(<R)$ as a function of the radius for different populations of the cluster sample. The solid/dashed lines represent the median values for the different subsamples, while the 1-$\sigma$ errors for the total sample are represented by the gray shaded region. The two horizzal dotted regions represent, from the top (in dark green) to the bottom (in grey), the 1-$\sigma$ confidence level on the primordial baryon fraction from {\em WMAP} ($\Omega_b/\Omega_m=0.167\pm0.006$) and {\em Planck} ($\Omega_b/\Omega_m=0.155\pm0.004$), respectively. The panel at the bottom represents the fractional variation of the profiles of the subsamples with respect to the whole sample. In order to improve the readability of the figures, we interpolated the data via a least-squares constrained spline approximation so as to have continuous functions. Note that $f_{\rm gas}(<R)$ scales with the Hubble constant as $f_{\rm gas}\propto h^{-3/2}$.}
\label{bias335wef2w}
\end{center}
\end{figure}

In Figure \ref{bias335wef2w} we present gas fraction profiles for the whole sample and for different subsamples. We then compare our measure gas fraction with the existing literature out to $R_{500}$. The value of $f_{\rm gas,500}=0.130\pm 0.002$ at $R_{500}$. We compare our gas fraction to previous studies of $f_{\textrm{gas}}$ using hydrostatic mass estimates. We used the results from the samples of relaxed clusters by \cite{vikhlinin2006} and \cite{ettori2009b}. The former measured $f_{\textrm{gas,500}}=0.110\pm{0.003}$ for 10 relaxed clusters observed with {\em Chandra} spanning a redshift range $z=0.02-0.23$; the latter $f_{\textrm{gas,500}}=0.106\pm{0.044}$ via 52 X-ray luminous galaxy clusters observed with {\em Chandra} in the redshift range $0.3-1.27$. These results are in marginal disagreement with our determination of the gas fraction at $R_{500}$. This stems mostly from the more recent {\em Chandra} calibrations adopted in the present work, which lead to a lower spectral temperature profile (by $\sim 10-15\%$) and hence to higher values of the gas fraction. Alternatively, our sample selection criteria could have some ramifications for the measurement of $f_{\textrm{gas}}$, since we are focusing on luminous and massive objects. Indeed the gas fraction is expected to be slightly larger for luminous objects \citep{landry2012}, while the aforementioned samples includes less massive systems as well. Our results are also in agreement with the determination of the gas fraction from \cite{eckert2013b} via \emph{ROSAT} X-ray and \emph{Planck} Sunyaev-Zeldovich (SZ) observations ($f_{\textrm{gas,500}}=0.132\pm0.005$).

With galaxy clusters at the crossroads of cosmology and astrophysics, here we discuss the implications of our findings for the use of the gas mass fraction as both cosmological and astrophysical probe.

\subsection{Implications for cosmological studies}\label{impl_cosm}

A number of studies have resolved the hot gas mass fraction from X-ray observations in galaxy clusters to place constraints on the cosmological parameters, in particular $\Omega_{\rm m}$, $\Omega_{\Lambda}$, and the ratio between the pressure and density of the dark energy, $w$ \citep{allen2008,ettori2009b}. This hot gas accounts for most of the baryons in clusters: the remaining baryons are in stars and intracluster light (ICL), and account for a few percent of the total mass \citep{lin2003}. The underlying idea of this cosmological test is that massive galaxy clusters are relatively well-isolated and gravity-dominated structures, where baryons and dark matter have accreted from very large regions of $\sim$10 comoving Mpc. 

Therefore, their relative baryon budget $f_{\rm gas}$ should be representative of the cosmic value $\Omega_b/\Omega_m=0.167\pm0.006$ predicted by the cosmic nucleosynthesis (light element formation during the Big Bang) and standard inflationary cosmology, as inferred from the {\em WMAP} 9-year data \citep{bennett2013}. Hence the baryon fraction can be a precise proxy of both the mean dark matter and dark energy density of the Universe.

Gas mass fraction measurements are complementary to (and competitive with) cosmological constraints using Type Ia supernovae (SNe Ia) and anisotropies of the cosmic microwave background \citep{allen2011}. However, they rely on the key assumptions that the hot gas fraction within a given radius is independent of redshift; can be converted to the universal fraction by correction factors which measure the ``baryon depletion`` from cosmological simulations and the fraction of baryons in stars and galaxies (not directly constrained via X-ray observations); and can be measured reliably.

In order to understand how biases on the hot gas fraction $f_{\rm gas}$, stellar fraction $f_{\textrm{stars}}$ and depletion parameter $b$ (the parameter used to convert the hot gas fraction to the cosmic ratio) propagate into the determination of, e.g. $\Omega_{\rm m}$, we can reverse-engineer the total matter content using the observed baryon fraction in clusters: 
\begin{equation}
\Omega_{\rm m}= b \; \Omega_{\textrm{b}}/(f_{\rm gas} + f_{\textrm{stars}}). 
\label{seeg2}
\end{equation}
with $f_{\textrm{stars}}\sim 0.12$ \citep{gonzalez2013}.

After correcting the hot gas fraction for the baryons in stars, our reconstructed average baryon fraction increases with radius and reaches the value $f_{\textrm{b},100}=0.177\pm0.005$, slightly larger than the primordial value. 

We point out that in our stacking analysis we assume a priori that the gas fraction, and hence the gas densities, do not depend on the system temperature (Equation \ref{dweded}). This hypothesis is verified a posteriori in light of a lack of dependency of the gas fraction on the cluster temperatures. This is probably explained by the relatively narrow temperature range spanned in our sample ($\gesssim$ 3 keV), such that the clusters in our sample should show little dependence of the physical properties on the global temperature. Indeed, while the gas fraction in groups and intermediate mass clusters implies an increasing trend of the cumulative gas mass fraction with temperature \citep{sun2009}, it is less clear whether such a trend persists for relatively massive objects. Measurements by \cite{vikhlinin2006} and \cite{allen2008} are both consistent with being constant with cluster temperature, for objects with $kT > 5$ keV, in agreement with our findings. Note that most of the clusters (264 out of 320) in our sample have temperatures greater than 5 keV. This trend is also confirmed by the level of the intrinsic scatter of $f_{\rm gas}$ ($\sim15\%$ at $R_{500}$, $\sim25\%$ at $R_{200}$), which does not appear to depend on the selected subsamples.

Our results also demonstrates that, when computed within the whole cluster virialization region, there is no relevant dependency of the gas fraction on cluster dynamical state at radii larger than $R_{500}$. These results are in tension with the findings of \cite{eckert2013b}, who finds significant differences between the baryon fraction of relaxed, cool-core (CC) systems and unrelaxed, non-cool core (NCC) clusters out to $R_{200}$. This difference might arise since they explicitly use hydrostatic equilibrium (by combining the pressure from SZ  {\em Planck} data with the gas density from {\em ROSAT}) to calculate total masses, which can bias upward the gas mass especially in unrelaxed systems \citep{battaglia2012,roncarelli2013}.

Appreciably high $f_{\rm gas}$ (larger than the cosmic value) towards the virial radius have been also reported in recent studies based on {\em Suzaku} data \citep[e.g.,][]{simionescu2011}, calling for a significant clumpy distribution of the ICM and the presence of non-thermal pressure support, in moderate disagreement with the present analysis. In this respect, understanding the unbiased value of the baryonic budget in clusters has clearly important ramifications in the use the hot gas fraction as a potential standard candle for the cosmology. For example, \cite{ettori2009b} found a tension between the inferred value of $\Omega_{\rm m}$ (jointly constrained with $\Omega_{\rm \Lambda}$) via the hot gas fraction in X-ray clusters ($\Omega_{\rm m} =  0.32^{+0.03}_{-0.02}$) and the value from {\em WMAP} 5-year data ($0.279 \pm 0.013$), with the former exceeding the latter. If the {\em WMAP} best-fit results are assumed to fix the cosmological parameters, this tension is expected to arise from biases due to non-thermal pressure support and ICM clumpiness.

We finally observe no evolution of the gas fraction with redshift. Measurements of the apparent lack of evolution of the cluster X-ray gas mass fraction has been used to probe the acceleration of the Universe, through the dependence of $f_{\rm gas}$ on the assumed distances to the clusters \citep{ettori2006,ettori2009b}. This assumption has never been tested observationally. Our results suggest, for the first time, that this assumption in tenable for cosmological purposes.

\subsection{Implications for the cluster astrophysics}\label{impl_astr}

The nearly closed-box nature of deep potentials of massive galaxy clusters makes them ideal laboratories to study non-gravitational processes, in particular galaxy feedback from supernovae and AGNs, operating during galaxy formation and their effects on the surrounding intergalactic medium. In the literature there has been a long-standing debate on the baryon content of clusters, where most of the studies indicate that the cluster baryon budget is typically lower than the cosmic baryon fraction \citep{vikhlinin2006,allen2008,ettori2009b}, although other analyses find that the baryon content is consistent with the cosmic ratio at $R_{500}$ \citep[][]{miller2012,landry2012}. A possible explanation of the observed deficit of baryons was that non-gravitational processes associated with galaxy formation, feedback from supernovae, AGN, star formation, or galactic winds may push the ICM towards the cluster outskirts \citep[$R\sim R_{200}$, see, e.g.][]{metzler1994,bialek2001,mccarthy2007,sun2009}.  Nevertheless, we stress that current studies are mostly limited to the inner volumes of the clusters ($R\lesssim R_{2500}-R_{500}$), because of the very low surface brightness of the X-ray signal in the outskirts.

As previously pointed out, we do not observe evolution of the gas fraction with redshift and an overall baryon budget slightly larger than the cosmological value within $R_{100}$. This might dispel questions which arose in the literature on the whereabouts of these putative {\em missing baryons}. This also confirms that the most massive clusters, which are relatively well-isolated and gravity-dominated structures, should retain all their gaseous matter. Moreover, while any feedback process will be more significant in the inner regions of low-mass systems when compared to the binding energy of the gas \citep{sun2009}, we checked that in the outskirts there seems to be no particular dependency on the cluster temperature or dynamical state. A posteriori, these findings suggest that our assumption of self-similarity holds for the ICM in the virialization region in clusters, regardless their dynamical states and mass.

We point out that an additional baryonic component might not be detected, raising the possibility that a non-negligible fraction of baryonic mass may be hidden in a form that is difficult to measure, such as cool gas. As we observed in \S\ref{compsim}, the presence of dense and cold clumps at temperatures $kT \lesssim 1$ keV in the ICM might remain undetected, since their exponential bremsstrahlung cutoff will be at energies $\sim kT/(1+z)$ below the energy threshold of the X-ray detector ($\sim 0.5$ keV for {\em Chandra}). Moreover, if the multiphase structure of the ICM turns out to be important as numerical simulations suggest \citep[see, e.g.,][]{rasia2014}, we might actually underestimate baryons due to the single-temperature modelling of the ICM (\S\ref{compsim}). In this respect, \cite{afshordi2007} stacked the SZ effect signals of {\em WMAP} observations of a large sample of massive clusters to constrain the thermal energy of the ICM. With the aid of hydrodynamic simulations, they converted this thermal energy into an estimate of the hot gas fraction of clusters, and found $f_{{\rm gas},200}=0.109\pm0.013$. This implies a deficit ($\sim35\%$) of baryons from the hot ICM (without correcting for $f_{\textrm{stars}}$), calling for a cool phase of the ICM. This effect could be competitive with the enhancement of the gas fraction due to inhomogeneities of the gas distribution ($\sim C^{0.5} \sim 1.2-1.4$, see \cite{roncarelli2013,battaglia2012} and below discussion).

We also observe that the gas faction along the direction of the filaments is systematically higher ($\sim10-15\%$) than the remaining volume. This picture is supported by our analysis of azimuthal variations of the gas density in cluster outskirts (\S\ref{lsst3}), which suggests that even CC clusters exhibit significant departures from spherical symmetry around $R_{200}$. Consequently, a full azimuthal coverage is required to study the baryon budget of cluster outer regions.

Next, we investigated the biases on our gas fraction measurements due to the effects of inhomogeneities of the gas distribution ("clumpiness") and non-thermal pressure. For the former bias, observations and simulations show that the gas clumping factor $C$ lies in the range $\sim 1.3-2$ at $R_{200}$ \citep{nagai2011,morandi2013b,morandi2014}, which will bias upwards the gas mass fraction of about $C^{0.5}\sim 1.2-1.4$ \citep{battaglia2012,roncarelli2013,morandi2013b,morandi2014}. For the latter bias, our measurements are also affected by violation of hydrostatic equilibrium (\S\ref{sys453}). A bias $b\sim 0.1$ translates into a bias (upwards) of $\sim 14\%$ of the gas fraction at $R_{200}$. We stress that the real challenge in future gas fraction measurements will be to account for these systematics, since nowadays X-ray studies customarily hinge on the assumptions of hydrostatic equilibrium, single-temperature spectral models and homogeneous distribution of the gas density. If these estimates of the biases due to non-thermal pressure and in particular clumpiness are correct, we might non-negligibly overestimate the gas fraction, leading to an actual deficit of baryons in clusters with respect to the cosmic value. However, this deficit might be dispelled if we change the cosmology from {\em WMAP} to {\em Planck}, as we discuss below.

Finally, we observe that the differences between the {\em WMAP} 9-year and our cosmological parameters affect the estimates of the gas mass fraction through the variation of the angular diameter distance $d_{\rm ang}$ ($f_{\rm gas}  \propto d_{\rm ang}(H_{0}, \Omega_{\rm m}, \Omega_{\Lambda}, w)^{3/2}$). If we assume the {\em WMAP} 9-year parameters, high-$z$ clusters will have a slightly larger ($\lesssim 2$ percent) gas fraction. 

Moreover, we analyzed the effects of changing the cosmology from {\em WMAP} to {\em Planck}. Plank data seem to suggest an higher value of the total matter density of the Universe ($\gesssim 0.3$) and a lower value of the Hubble constant with respect to {\em WMAP} \citep{planck2013b}. This translates into a lower value of the cosmic baryon fraction $\Omega_{\textrm{b}} / \Omega_{\rm m} = 0.155\pm0.003$. On the other hand, the net effect of switching from {\em WMAP} to {\em Planck} cosmology is to increase the measured gas fractions by $(d_{\rm ang,Plank}/d_{\rm ang,WMAP})^{3/2}\sim 1.05$  for high-$z$ systems, and stellar fractions by $(d_{\rm ang,Plank}/d_{\rm ang,WMAP})\sim 1.03$. Clearly, the most striking effect of the use of the {\em Planck} cosmological parameters is to increase the baryon content within $R_{100}$ \citep[see also][]{gonzalez2013}.

\section{Summary and conclusions}\label{conclusion33}
Cluster outskirts are the regions where the transition between virialized cluster gas and accreting material from the large-scale structure occurs. In this paper, we have presented our analysis of a sample of 320 clusters ($z=0.056-1.24$) from the {\em Chandra} archive, focusing on the properties of the gas in cluster outskirts.

We have exploited the large archival dataset of {\em Chandra} clusters to trace the gas density and gas fraction of the intracluster gas out to $R_{100}$. We measured the median of the distribution of the (renormalized) emission measure profiles $EM(R/R_{200})$ to detect signal in the outer volumes and measured the typical gas density, gas slope and gas fraction. We investigated the evolution of the physical properties as a function of dynamical state, redshift, cluster temperatures, as well as their azimuthal variation. Finally, we compared our average density and gas fraction with the results of numerical simulations.

In tension with some recent {\em Suzaku} results, and confirming previous evidence from {\em ROSAT} and {\em Chandra}, we observe a steepening of the density profiles beyond $R_{500}$. We report that galaxy clusters deviate significantly from spherical symmetry, but with negligible differences between relaxed and disturbed systems. Thus, our findings on different subsamples suggest that our assumption of self-similarity holds for the ICM in the virialization region in clusters, regardless their dynamical states, temperature and redshift.

Comparing our results with numerical simulations, we find that the latter simulations fail to reproduce the gas distribution, even well outside cluster cores. Simulations including cooling and star formation convert a large amount of gas into stars, which results in a low gas density and gas mass fraction with respect to the observations.

We then measured the hot gas fraction in galaxy clusters observed with {\em Chandra} out to $R_{100}$. After converting the hot gas fraction to the total baryon budget in clusters, we measured a value of the baryon fraction which slightly exceeds the cosmic baryon fraction. However, a careful analysis of the systematics due to non-thermal pressure and clumpiness suggest that we might non-negligibly overestimate the total baryon budget, leading to an actual deficit of baryons in clusters with respect to the cosmic value.

This study has important implications on our knowledge of the state of the gas in cluster outskirts, and promises to help advance cosmological studies based on the cluster growth over cosmic time in the coming decade.

\section{Future works}\label{conclusion3563}
Our work opened the way to a novel method, the stacking the X-ray signal of cluster outskirts, exploiting the high level of self-similarity of the cluster outer volumes. The total exposure time is $\sim 20$~Ms for the whole sample, which is one order of magnitude larger than the deepest {\em Chandra} observations (e.g. compared to the 2 Ms of A133, see Vikhlinin et al. in prep.), and 2-3 order of magnitudes larger than an average X-ray observation. Clearly, the most striking ramification of this analysis is the ability to greatly exploit the {\em Chandra} database, while no single observation could achieve the same scientific goals.

In future papers will use the stacking approach to infer constraints on the cosmological parameters via the analysis of the X-ray in the outskirts (Morandi et al. in prep.). This novel approach will pin down the cosmological parameters, in particular $\Omega_{\rm m}$ and $w$ (the ratio between the pressure and density of the dark energy), with a greatly improved accuracy with respect to the current analyses of clusters, characterized by  relatively smaller samples and limited to the interiors of the sources (e.g. out to $R_{2500},R_{500}$). We plan to use the X-ray stacked data to improve further our understanding of how the integrated Compton parameter scales with total cluster mass and the corresponding Compton SZ parameter from {\em Planck}, when available; and to constrain the temperature and the metallicity profiles in the outer regions, in order to understand the thermodynamic of the outer regions.



\section*{acknowledgements}
We are indebted to  Erwin Lau, Stefano Ettori, Mauro Roncarelli, Norbert Werner, Daisuke Nagai, Craig Sarazin, Ben Maughan, Dominique Eckert and Alexey Vikhlinin for valuable comments. We thank Terry Gaetz for the work and information on the ACIS stowed background. We thank the anonymous referee for the careful reading of the manuscript and suggestions, which improved the presentation of our work. A.M. gratefully acknowledges the hospitality of the Harvard-Smithsonian Center for Astrophysics. A.M. and M.S. acknowledge support from Chandra grants GO2-13160A, GO2-13102A, and NASA grant NNX14AI29G.

\begin{appendix}
 
\section{Distribution of the X-ray counts in the sample}\label{sample4634}

\begin{figure*}
\begin{center}
\hbox{
\psfig{figure=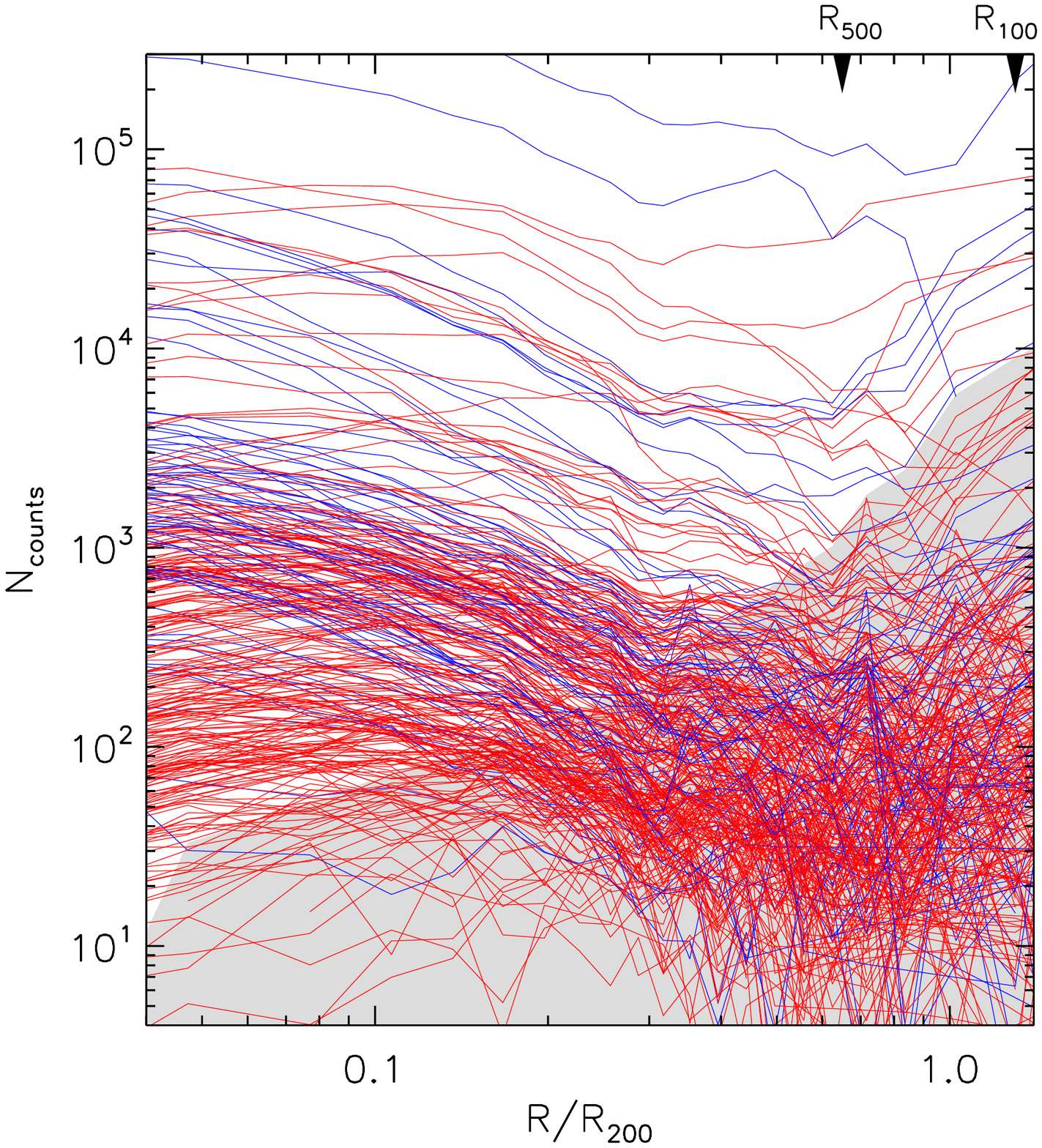,width=0.5\textwidth}
\psfig{figure=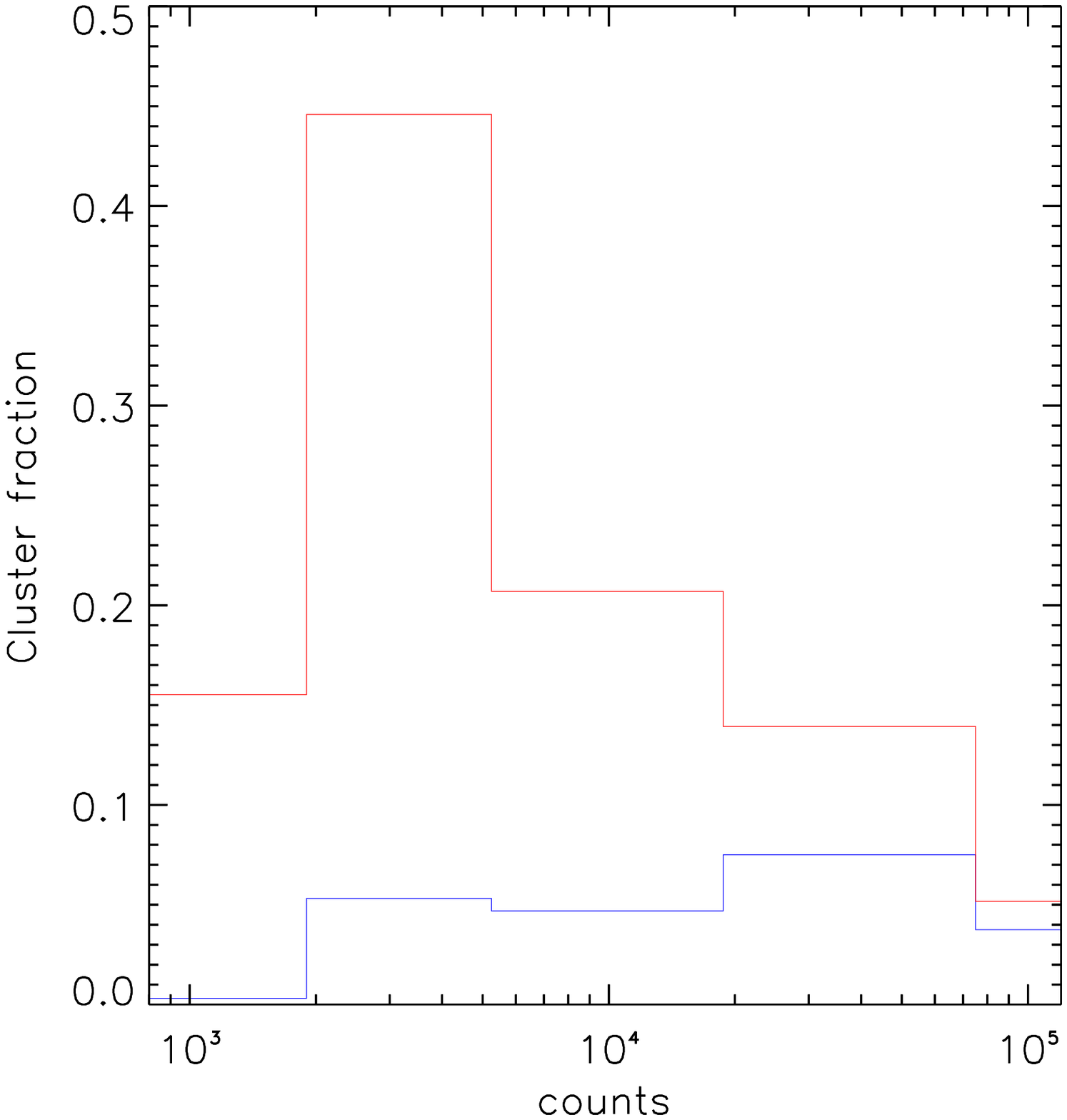,width=0.5\textwidth}
}
\caption{Left panel: number of net X-ray counts per annulus in the energy range 0.7-2.0 keV for the 320 clusters of our sample. The shaded gray band represents the measured total background in each annulus. Central panel: distribution of total net X-ray counts for the 320 clusters used in this work. For all the panels blue and red curves (dots) correspond to clusters classified as cool core and non-cool core, respectively (see \S\ref{morph}).}
\label{sample4634f}
\end{center}
\end{figure*}
We present the number of net X-ray counts per annulus for our sample. We also show the distribution of the number of net X-ray counts for whole dataset.

\section{Spectral temperatures dependency on atomic database and absorption}\label{app1}

\begin{figure*}
\begin{center}
\hbox{
\psfig{figure=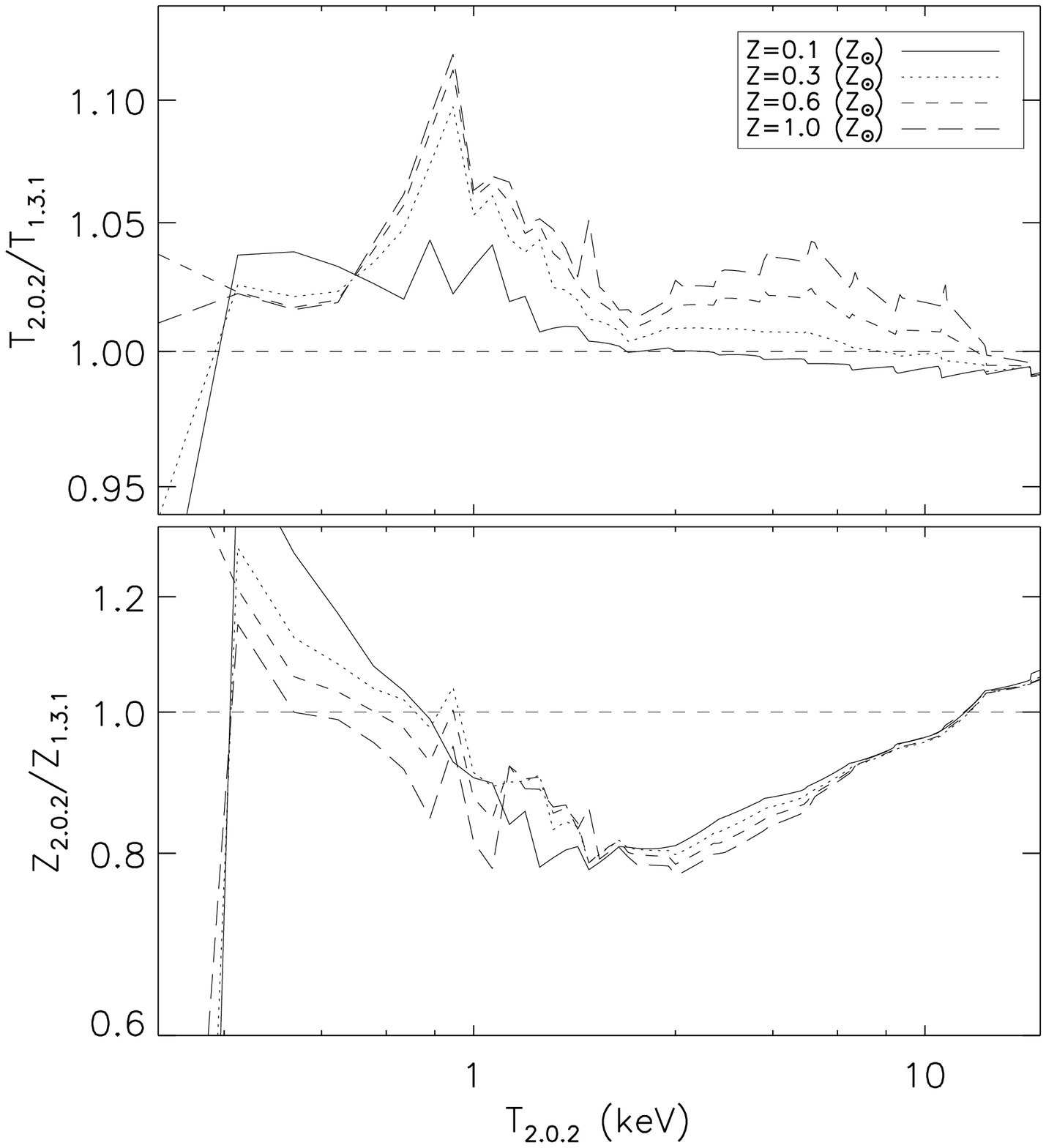,width=0.5\textwidth}
\psfig{figure=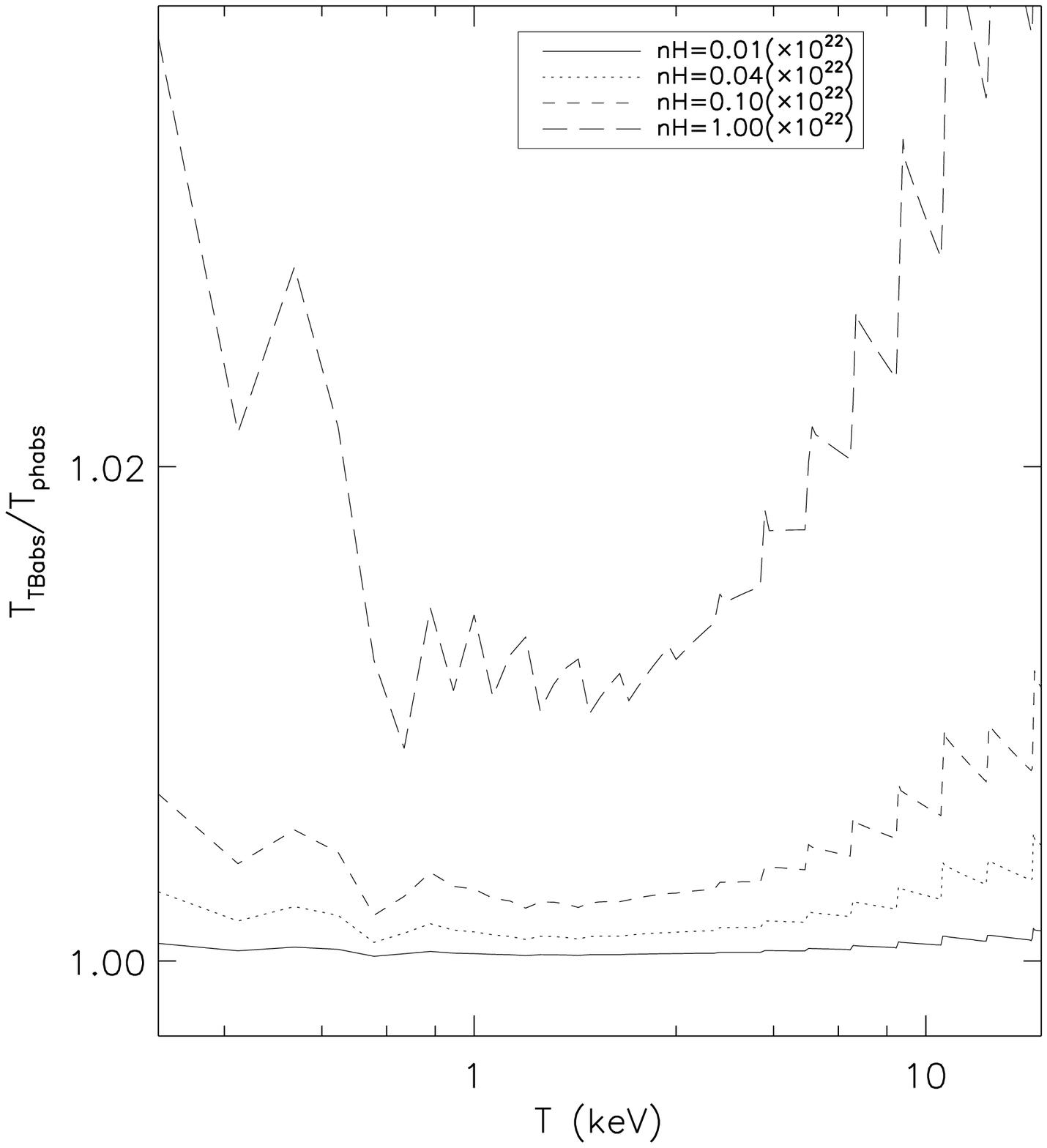,width=0.5\textwidth}
}
\caption{Left panel: comparison between the spectral temperatures and metallicities recovered via the AtomDB databases 2.0.2 and 1.3.1 via mock data. Right panel: comparison between the spectral temperatures recovered via different models of X-ray absorption by the ISM via mock data (\texttt{tbabs} v.s. \texttt{phabs}).}
\label{evderyddd2}
\end{center}
\end{figure*}

We present a comparison between spectral temperatures and metallicities recovered via the AtomDB databases 2.0.2 and 1.3.1, and via different models of X-ray absorption by the ISM (Figure \ref{evderyddd2}). Note that in our work we adopted the APEC emissivity model \citep{foster2012} and the AtomDB (version 2.0.2) database of atomic data, and we employed the solar abundance ratios from \cite{asplund2009}. We also used the Tuebingen-Boulder absorption model (\texttt{tbabs}) for X-ray absorption by the ISM. 

Finally, we discuss the bias due to our choice to fix the hydrogen column density $N_H$ to the Galactic value by using the Leiden/Argentine/Bonn (LAB) HI-survey \citep{kalberla2005}. This radio surveys measure the hydrogen column density from the HI-21 cm line, only providing the neutral hydrogen along the line of sight, while the molecular and ionized hydrogen is not accounted for. \cite{willingale2013} provided a method to account for the molecular hydrogen via the dust extinction measured in the B and V bands, and by using X-ray afterglows of gamma ray bursts. Although the neutral hydrogen usually contributes most of the total hydrogen for $N_H\lesssim 10^{21} {\mbox{cm}}^{-2}$, the impact of molecular hydrogen should be important for larger column densities. For each cluster, we performed mock simulations of absorbed spectra by assuming the our average gas density profile (\S\ref{sys453b}), the average temperature profile as measured by \cite{vikhlinin2006} and hydrogen column density values including both neutral (from the aforementioned radio surveys) and molecular hydrogen \citep[as determined by][]{willingale2013}. These mock spectra have been then fit by fixing the hydrogen column density to the Galactic value as determined by the HI-survey \citep{kalberla2005}. For $N_H\lesssim 6\times 10^{20}{\mbox{cm}}^{-2}$, the bias (upwards) on the recovered spectral temperature is $\lesssim 1-2\%$, which translates into a bias of similar magnitude (downwards) on the gas density $n_{e,200}$. However, for $N_H> 10^{21}{\mbox{cm}}^{-2}$ the bias is appreciable ($\gesssim 10\%$). Note that most of the clusters (290 out of 320) have column densities smaller than $6\times 10^{20}{\mbox{cm}}^{-2}$, while only 3 clusters have $N_H> 10^{21}{\mbox{cm}}^{-2}$. Hence the overall systematics on the gas density recovered via stacking is estimated $\lesssim 2\%$ at $R_{200}$.

\section{Definition of cooling core systems}\label{maugh1ehrhgr}
\begin{figure}
\begin{center}
\psfig{figure=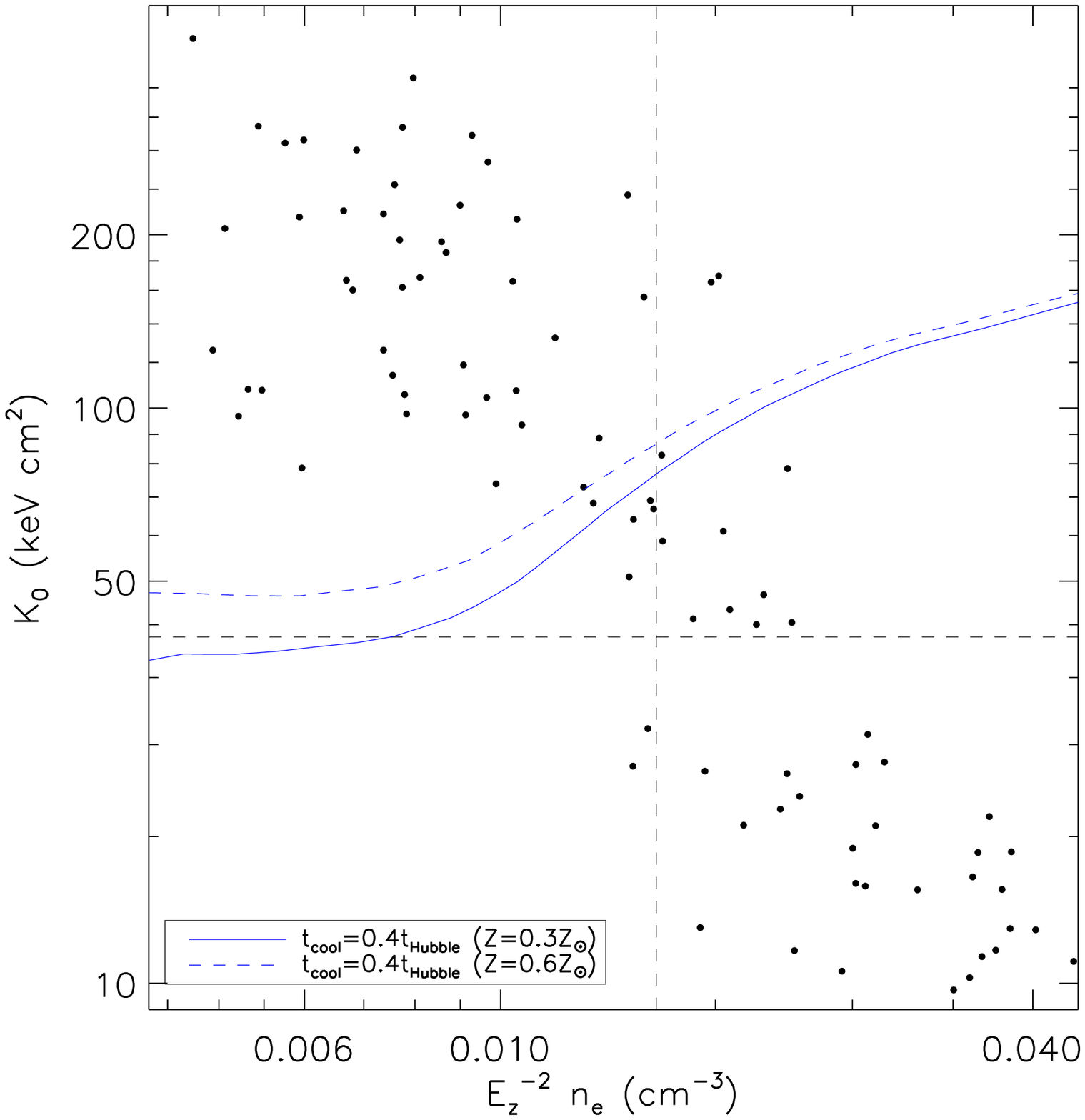,width=0.44\textwidth}
\caption{Comparison between the central entropy $K_0$ from the subsample of 93 objects in common with \citet{cavagnolo2009} with our gas density at $0.03R_{500}$. The threshold we use to define a cool core systems is $E_z^{-2} n_{e,0}>1.5\times 10^{-2}{\mbox{cm}}^{-3}$, while \citet{cavagnolo2009} use the entropy threshold $K_0\sim 40 \, \mbox{keV cm}^{2}$ to approximately demarcate the division between CC (cool core clusters) and NCC (non-cool core clusters). The blue line divide the plot in two region with cooling time greater (upper region) and smaller (lower region) than 0.4 Hubble times.}
\label{evderyddrthtyd245}
\end{center}
\end{figure}

We classify clusters as cool cores on the basis of density $E_z^{-2} n_{e,0}$ at $0.03\,R_{500}$. The threshold we use to define a cool core systems is $E_z^{-2} n_{e,0}>1.5\times 10^{-2}{\mbox{cm}}^{-3}$. We compared our definition of cool core systems with \cite{cavagnolo2009}, whose sample shares 93 objects in common with ours. They use the entropy threshold $K_0=30-50 \, \mbox{keV cm}^{2}$ to approximately demarcate the division between CC and NCC. In Figure \ref{evderyddrthtyd245} we present a comparison between the two definitions of cooling core cluster, along with the definition of cool core based on the cooling time. As shown in Figure \ref{evderyddrthtyd245}, our definition of cool cores is roughly agrees with the definition of \cite{cavagnolo2009}, while the definition based on cooling time falls in between. In the end, our conclusions on the CC/NCC classification are not sensitive to our definition of cool cores.

\section{Comparison of $M_{gas,500}$ with the literature}\label{maugh1}

\begin{figure}
\begin{center}
\psfig{figure=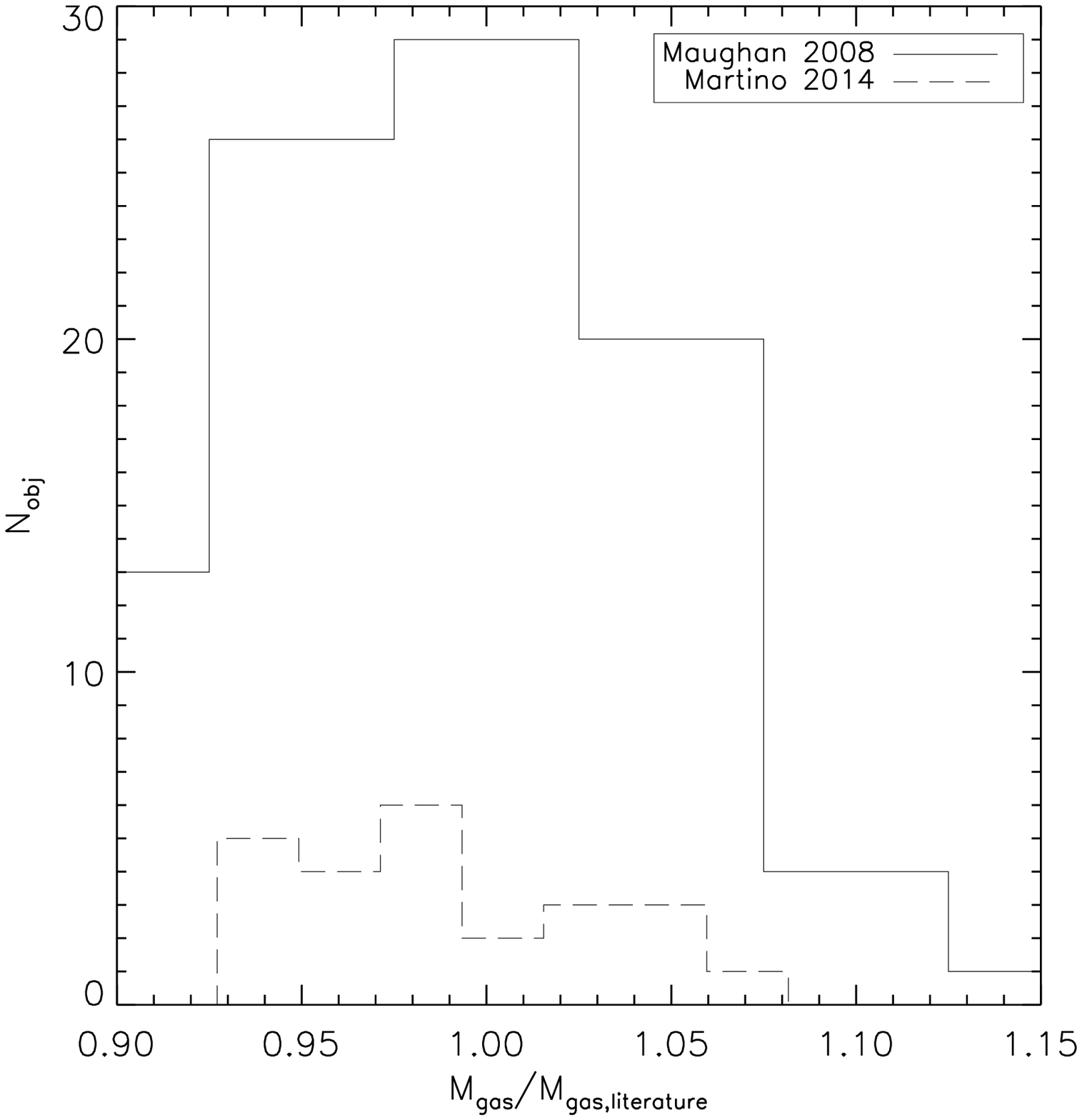,width=0.44\textwidth}
\caption{Comparison between of the cumulative gas mass $M_{gas,500}$ at $R_{500}$ with the subsample of 95 objects in common with \citet{maughan2008} and the 32 objects in common with \citet{martino2014}. The comparison has been performed at the same $R_{500}$ from the literature, and the gas mass has been recovered from individual gas density profiles rather than the stacked profiles.}
\label{evderyddd2ffff}
\end{center}
\end{figure}

We compared our gas mass at $R_{500}$ (recovered via deprojection of individual emission measure profiles) with the results of \cite{maughan2008}, whose sample shares 95 objects in common with ours, and with \cite{martino2014}, whose sample contains 32 objects in common with us (Figure \ref{evderyddd2ffff}). We found no significant differences between the recovered gas masses (\S\ref{maugh1}), with a scatter of $\sim5\%$. We performed the comparison at the same $R_{500}$ from the literature, and the gas mass has been recovered from individual gas density profile rather than the stacked profiles. 

\end{appendix}






\newcommand{\noopsort}[1]{}

\clearpage

\begin{table*}
\begin{center}
\caption{The X-ray properties of the galaxy clusters in our sample. For each object different columns report the cluster name, the redshift $z$, the exposure time $t_{\rm exp}$,  the identification number of the {\it Chandra} observations,  the spectroscopic temperature $T_{\rm ew}$ in the radial range $0.15-0.75\; R_{500}$, the centroid shift (in units of $R_{500}$), and a flag for the presence or not of a cooling core (labeled CC and NCC, respectively). }
\begin{tabular}{l@{\hspace{0.8em}} c@{\hspace{0.8em}} c@{\hspace{0.8em}} c@{\hspace{0.8em}} c@{\hspace{0.8em}} c@{\hspace{0.8em}} c@{\hspace{0.8em}} c@{\hspace{0.8em}} c@{\hspace{0.8em}} c@{\hspace{0.8em}} c@{\hspace{0.8em}} c@{\hspace{0.8em}} c@{\hspace{0.8em}} }
\hline \\
Cluster   & $ z $ & $ t_{\rm exp} $ & ID &   $T_{\rm ew}$ & $w$ & CC/NCC\\
          &        &        ks         & &  (keV)        & ($\times 10^{-2}$)  &\\
\hline \\
              A85&   0.056 &   272.4&    904  4881  4882  4883  4884  4885  4886  4887& $   6.45\pm   0.86$ & $   0.49\pm   0.18$ &     CC \\
 &  & &    4888 15173 16264 15174 16263&  &\\
              A133&   0.057 &  2393.0&  13442 13443 13444 13445 13446 13447 13448 13449& $   3.76\pm   0.25$ & $   0.26\pm   0.09$ &     CC \\
 &  & &   13450 13451 13452 13453 13454 13455 13456 13457&  &\\
 &  & &   14333 14338 14343 14345 14346 14347 14354 13391&  &\\
 &  & &   13392 13518  3183  3710 12177 12178 12179  9897&  &\\
              A644&   0.070 &    48.9&   2211 10420 10421 10422 10423& $   8.12\pm   1.12$ & $   1.56\pm   0.57$ &     CC \\
              A401&   0.075 &   153.9&  10416 10417 10418 10419 14024& $   5.44\pm   0.86$ & $   1.03\pm   0.38$ &    NCC \\
             A2029&   0.076 &    29.2&   6101 10434 10435 10436 10437& $   7.38\pm   0.92$ & $   0.21\pm   0.08$ &     CC \\
             A1650&   0.084 &   218.4&   7691  5822  5823  6356  6357  6358  7242 10424& $   5.89\pm   0.86$ & $   0.38\pm   0.14$ &     CC \\
 &  & &   10425 10426 10427&  &\\
             A1068&   0.137 &    19.5&  13595 13596 13597 13598& $   6.38\pm   1.26$ & $   0.29\pm   0.11$ &     CC \\
             A2276&   0.141 &    39.5&  10411& $   3.40\pm   0.48$ & $   0.49\pm   0.18$ &    NCC \\
             A1413&   0.143 &   155.6&   5003 12194 12195 12196 13128  1661  5002   537& $   7.78\pm   0.22$ & $   0.57\pm   0.21$ &     CC \\
 &  & &    7696&  &\\
             A3402&   0.146 &    19.8&  12267& $   3.09\pm   0.55$ & $   1.39\pm   0.51$ &    NCC \\
             A2409&   0.148 &    10.2&   3247& $   5.40\pm   0.40$ & $   1.09\pm   0.40$ &    NCC \\
   RXCJ0449.9-4440&   0.150 &    19.8&   9417& $   4.23\pm   0.47$ & $   0.66\pm   0.24$ &    NCC \\
             A2204&   0.152 &   106.5&   7940  6104 12895 12896 12897 12898& $  10.38\pm   0.14$ & $   0.18\pm   0.06$ &     CC \\
              A907&   0.153 &    99.0&   3185  3205   535& $   6.60\pm   0.24$ & $   0.17\pm   0.06$ &     CC \\
    RXJ1000.4+4409&   0.154 &    18.5&   9421& $   3.42\pm   0.41$ & $   3.42\pm   1.25$ &     CC \\
             A3866&   0.154 &     9.0&  15162& $   6.01\pm   0.84$ & $   0.15\pm   0.05$ &     CC \\
            Zw8284&   0.156 &    22.0&  15118& $   5.00\pm   0.46$ & $   0.50\pm   0.18$ &    NCC \\
   RXCJ0528.2-2942&   0.158 &    19.3&   9418& $   6.41\pm   0.48$ & $   0.75\pm   0.27$ &    NCC \\
             A1240&   0.159 &    51.4&   4961& $   6.85\pm   0.85$ & $   2.05\pm   0.75$ &    NCC \\
             A2259&   0.164 &    10.0&   3245& $   6.37\pm   0.44$ & $   0.60\pm   0.22$ &    NCC \\
             A2445&   0.165 &    26.7&  12249& $   5.62\pm   0.77$ & $   0.63\pm   0.23$ &    NCC \\
              A853&   0.166 &    24.7&  12250& $   6.50\pm   1.27$ & $   0.52\pm   0.19$ &    NCC \\
             A1201&   0.169 &    47.4&   9616& $   6.47\pm   0.41$ & $   2.54\pm   0.93$ &    NCC \\
              Z808&   0.169 &    18.8&  12253& $   6.95\pm   1.65$ & $   0.24\pm   0.09$ &     CC \\
            ZwCl15&   0.170 &    26.7&  12251& $   6.11\pm   0.63$ & $   0.43\pm   0.16$ &    NCC \\
             A1204&   0.171 &    23.3&   2205& $   4.32\pm   0.39$ & $   0.22\pm   0.08$ &     CC \\
             A1914&   0.171 &    38.6&   3593 12197 12892 12893 12894& $   9.18\pm   0.91$ & $   0.64\pm   0.23$ &     CC \\
              A586&   0.171 &     9.9&  11723& $   6.85\pm   1.01$ & $   0.48\pm   0.18$ &     CC \\
   RXCJ0616.3-2156&   0.171 &    23.8&  15100& $   6.73\pm   0.35$ & $   2.66\pm   0.97$ &    NCC \\
    RXJ1750.2+3505&   0.171 &    19.8&  12252& $   6.14\pm   1.01$ & $   0.40\pm   0.15$ &     CC \\
             A3140&   0.173 &    20.0&   9416& $   6.25\pm   0.81$ & $   0.47\pm   0.17$ &    NCC \\
             A2218&   0.176 &    46.3&   7698  1666& $   7.96\pm   0.45$ & $   1.00\pm   0.36$ &    NCC \\
     MS0906.5+1110&   0.180 &    34.5&    924  7699& $   5.57\pm   0.44$ & $   0.99\pm   0.36$ &    NCC \\
              A665&   0.182 &   102.3&  12286 13201 15147 15148& $   8.05\pm   0.23$ & $   4.15\pm   1.51$ &    NCC \\
             A2187&   0.183 &    18.3&   9422& $   7.55\pm   0.74$ & $   1.61\pm   0.59$ &    NCC \\
             A1689&   0.183 &   196.7&   6930  7289  7701   540  1663  5004& $  10.27\pm   0.26$ & $   0.52\pm   0.19$ &     CC \\
             A0598&   0.186 &    19.8&  10442& $   6.28\pm   1.39$ & $   0.78\pm   0.29$ &     CC \\
              A383&   0.187 &    29.2&   2320   524& $   5.47\pm   0.32$ & $   0.20\pm   0.07$ &     CC \\
   RXCJ0331.1-2100&   0.188 &    20.0&  10790  9415& $   6.03\pm   0.73$ & $   0.80\pm   0.29$ &     CC \\
  CIZAJ2242.8+5301&   0.192 &   195.9&  14019 14020& $   9.37\pm   0.39$ & $   2.99\pm   1.09$ &    NCC \\
             A4023&   0.193 &    22.8&  15124& $   7.91\pm   1.12$ & $   2.56\pm   0.94$ &    NCC \\
     MS0839.8+2938&   0.194 &     4.8&   7702& $   4.49\pm   0.50$ & $   0.91\pm   0.33$ &     CC \\
              A115&   0.195 &   360.2&  13458 13459 15578 15581  3233& $   7.85\pm   0.22$ & $   4.12\pm   1.50$ &     CC \\
             A2507&   0.196 &    33.3&  12248& $   9.86\pm   1.62$ & $   0.53\pm   0.19$ &    NCC \\
      RXJ2247+0337&   0.200 &    49.0&    911& $   3.04\pm   0.47$ & $   1.46\pm   0.53$ &    NCC \\
             A3322&   0.200 &    15.6&  15111& $   7.47\pm   1.05$ & $   0.54\pm   0.20$ &    NCC \\
              A520&   0.202 &   408.1&   9424  4215  9426  9430   528& $   8.74\pm   0.25$ & $   1.88\pm   0.68$ &    NCC \\
             A3399&   0.203 &    24.5&  15125& $   6.76\pm   0.76$ & $   1.41\pm   0.51$ &    NCC \\
   ZwCl1829.3+6912&   0.204 &    64.0&  10412 10931& $   4.85\pm   0.64$ & $   0.83\pm   0.30$ &    NCC \\
              A963&   0.206 &     5.1&   7704& $   7.40\pm   1.75$ & $   0.38\pm   0.14$ &     CC \\
              A209&   0.206 &    20.0&    522  3579& $   8.69\pm   0.64$ & $   0.76\pm   0.28$ &    NCC \\
              A223&   0.207 &    44.8&   4967& $   4.16\pm   0.43$ & $   1.28\pm   0.47$ &    NCC \\
      RXJ0439+0520&   0.208 &    38.1&    527  9369  9761& $   5.14\pm   0.55$ & $   0.38\pm   0.14$ &     CC \\

\end{tabular}

 
\label{tab:1b}
\end{center}
\end{table*}

\addtocounter{table}{-1}
\begin{table*}
\begin{center}
\caption{Continued.}
\begin{tabular}{l@{\hspace{0.1em}} c@{\hspace{.5em}} c@{\hspace{0.8em}} c@{\hspace{0.8em}} c@{\hspace{0.8em}} c@{\hspace{0.8em}} c@{\hspace{0.8em}} c@{\hspace{0.8em}} c@{\hspace{0.8em}} c@{\hspace{0.8em}} c@{\hspace{0.8em}} c@{\hspace{0.8em}} c@{\hspace{0.8em}} }
\hline \\
Cluster   & $ z $ & $ t_{\rm exp} $ & ID &   $T_{\rm ew}$ & $w$ & CC/NCC\\
          &        &        ks         & &  (keV)        & ($\times 10^{-2}$)  &\\
\hline \\
  MACSJ0547.0-3904&   0.210 &    21.5&   3273& $   4.08\pm   0.39$ & $   1.77\pm   0.65$ &     CC \\
          ZWCL2701&   0.210 &     5.1&   7706& $   4.70\pm   0.62$ & $   1.07\pm   0.39$ &     CC \\
     G286.58-31.25&   0.210 &    21.9&  15115& $   7.03\pm   0.95$ & $   1.42\pm   0.52$ &    NCC \\
             A1430&   0.211 &    21.8&  15119& $   6.74\pm   0.73$ & $   2.06\pm   0.75$ &    NCC \\
              A222&   0.213 &    45.1&   4967& $   3.88\pm   0.30$ & $   3.65\pm   1.33$ &    NCC \\
             A1246&   0.213 &     5.0&  11770& $   6.03\pm   0.99$ & $   1.08\pm   0.40$ &    NCC \\
             A1423&   0.214 &    35.5&    538 11724& $   6.61\pm   0.53$ & $   0.67\pm   0.24$ &     CC \\
     MS0735.6+7421&   0.216 &   475.3&  10468 10469 10470 10471 10822 10918 10922& $   6.64\pm   0.14$ & $   0.47\pm   0.17$ &     CC \\
              A773&   0.217 &    60.2&   5006  3588   533 13591 13592 13593 13594& $   7.53\pm   0.49$ & $   1.05\pm   0.38$ &    NCC \\
   RXCJ1947.3-7623&   0.217 &    19.8&  15102& $   7.16\pm   1.15$ & $   0.83\pm   0.30$ &     CC \\
             A3084&   0.219 &    19.8&   9413& $   6.46\pm   0.80$ & $   0.78\pm   0.29$ &    NCC \\
             A3017&   0.220 &    14.9&  15110& $   7.11\pm   1.14$ & $   1.61\pm   0.59$ &     CC \\
   RXCJ0510.7-0801&   0.220 &    20.7&  14011& $   6.44\pm   0.43$ & $   1.71\pm   0.63$ &    NCC \\
              A368&   0.220 &    18.4&   9412& $   8.28\pm   1.25$ & $   0.49\pm   0.18$ &     CC \\
     MS1006.0+1202&   0.221 &    67.5&    925 13390& $   6.30\pm   0.48$ & $   2.16\pm   0.79$ &    NCC \\
            AS0592&   0.222 &    19.9&   9420& $   8.91\pm   1.08$ & $   0.72\pm   0.26$ &     CC \\
   RXCJ1514.9-1523&   0.223 &    59.0&  15175& $   9.41\pm   0.69$ & $   1.83\pm   0.67$ &    NCC \\
             A1763&   0.223 &    19.6&   3591& $   8.63\pm   0.82$ & $   1.16\pm   0.42$ &    NCC \\
             A1942&   0.224 &    60.6&   3290  7707& $   4.92\pm   0.46$ & $   1.14\pm   0.42$ &    NCC \\
             A2261&   0.224 &    24.3&   5007& $   8.09\pm   0.62$ & $   0.28\pm   0.10$ &     CC \\
             A1895&   0.225 &    18.7&  15129& $   4.93\pm   0.84$ & $   1.26\pm   0.46$ &    NCC \\
             A2219&   0.226 &   152.5&  13988 14355 14356 14431 14451  7892& $  10.81\pm   0.31$ & $   1.23\pm   0.45$ &    NCC \\
   ZwCl0823.2+0425&   0.226 &    21.4&  10441& $   4.83\pm   0.51$ & $   0.57\pm   0.21$ &    NCC \\
   RXCJ0118.1-2658&   0.228 &    19.7&   9429& $   7.45\pm   0.57$ & $   2.38\pm   0.87$ &    NCC \\
    RXJ0220.9-3829&   0.229 &    19.9&   9411& $   5.10\pm   0.65$ & $   0.37\pm   0.13$ &     CC \\
    RXJ1234.2+0947&   0.229 &    29.3&  11727   539& $   6.30\pm   1.01$ & $   1.99\pm   0.73$ &    NCC \\
              A141&   0.230 &    19.9&   9410& $   6.01\pm   0.81$ & $   1.24\pm   0.45$ &    NCC \\
              A267&   0.230 &    19.9&   3580& $   7.92\pm   0.62$ & $   1.46\pm   0.53$ &    NCC \\
             A2111&   0.230 &    31.2&    544 11726& $   8.66\pm   0.75$ & $   2.51\pm   0.92$ &    NCC \\
             A2667&   0.230 &    19.4&  13599 13600 13601 13602& $   6.31\pm   1.05$ & $   0.34\pm   0.13$ &    CC \\
    RXJ0439.0+0715&   0.230 &    18.8&   3583& $   6.30\pm   0.62$ & $   0.90\pm   0.33$ &     CC \\
             A1682&   0.234 &    29.4&   3244 11725& $   6.47\pm   0.70$ & $   0.85\pm   0.31$ &    NCC \\
             A2146&   0.234 &   375.5&  12245 13120 12246 13138 12247 13023 13020 13021& $   7.28\pm   0.20$ & $   3.63\pm   1.32$ &     CC \\
          ZwCl2089&   0.235 &     9.1&   7897& $   5.13\pm   0.71$ & $   0.23\pm   0.08$ &     CC \\
    RXJ2129.6+0005&   0.235 &    39.6&    552  9370& $   7.78\pm   0.40$ & $   0.71\pm   0.26$ &     CC \\
             A2465&   0.245 &    69.2&  14010 15547& $   3.97\pm   0.39$ & $   1.53\pm   0.56$ &    NCC \\
             A2125&   0.247 &   117.3&   2207  6891  7708& $   3.56\pm   0.25$ & $   2.53\pm   0.92$ &    NCC \\
             A2485&   0.248 &    19.8&  10439& $   5.86\pm   0.53$ & $   0.35\pm   0.13$ &    NCC \\
             A2645&   0.251 &    19.0&  14013& $   6.35\pm   1.01$ & $   0.83\pm   0.30$ &    NCC \\
             A1835&   0.253 &   192.4&   6880  6881  7370& $  10.00\pm   0.29$ & $   0.35\pm   0.13$ &     CC \\
              A521&   0.253 &    88.4&  12880 13190& $   7.51\pm   0.41$ & $   5.95\pm   2.17$ &    NCC \\
             A3088&   0.253 &    18.9&   9414& $   8.22\pm   1.04$ & $   0.28\pm   0.10$ &     CC \\
               A68&   0.255 &    10.0&   3250& $   8.97\pm   1.91$ & $   1.14\pm   0.42$ &    NCC \\
     MS1455.0+2232&   0.259 &   108.0&   4192  7709   543& $   5.26\pm   0.22$ & $   0.32\pm   0.12$ &     CC \\
     RXCJ1327+0211&   0.262 &     6.7&  11771& $   8.76\pm   1.70$ & $   1.10\pm   0.40$ &    NCC \\
    RXJ0528.9-3927&   0.263 &   109.9&   4994 15658 15177& $  10.64\pm   0.63$ & $   1.59\pm   0.58$ &     CC \\
             Z5768&   0.266 &    37.3&  14014  7898& $   3.29\pm   0.40$ & $   3.16\pm   1.15$ &    NCC \\
  MACSJ2211.7-0349&   0.270 &    17.7&   3284& $  11.97\pm   2.69$ & $   0.86\pm   0.32$ &     CC \\
   RXCJ0532.9-3701&   0.271 &    24.8&  15112& $   9.20\pm   1.24$ & $   0.40\pm   0.15$ &     CC \\
   RXCJ0303.7-7752&   0.274 &    33.6&  15113& $   9.83\pm   0.74$ & $   1.37\pm   0.50$ &    NCC \\
    SDSSJ1233+1511&   0.275 &    12.1&  11761& $   5.17\pm   0.61$ & $   3.46\pm   1.26$ &    NCC \\
             A1622&   0.275 &    12.9&  11763& $   4.76\pm   0.87$ & $   1.24\pm   0.45$ &    NCC \\
    RXJ0142.0+2131&   0.277 &    19.4&  10440& $   8.50\pm   0.66$ & $   1.21\pm   0.44$ &    NCC \\
             A2631&   0.278 &    26.0&   3248 11728& $   7.15\pm   0.82$ & $   1.84\pm   0.67$ &    NCC \\
     ACTJ0235-5121&   0.278 &    19.8&  12262& $   8.18\pm   0.76$ & $   2.28\pm   0.83$ &    NCC \\
    SDSSJ0104+0003&   0.278 &    15.0&  11765& $   6.49\pm   0.80$ & $   2.83\pm   1.03$ &    NCC \\
             A1576&   0.279 &    43.4&   7938 15127& $   7.14\pm   0.83$ & $   1.43\pm   0.52$ &    NCC \\
    RXJ2011.3-5725&   0.279 &    24.0&   4995& $   4.05\pm   0.49$ & $   0.73\pm   0.27$ &     CC \\
             A1758&   0.279 &   154.7&   7710 13997 15538 15540& $   8.64\pm   0.57$ & $   1.36\pm   0.50$ &    NCC \\
              A697&   0.282 &    19.5&   4217& $  11.20\pm   1.49$ & $   0.31\pm   0.11$ &    NCC \\ 
\end{tabular}
\end{center}
\end{table*}

\addtocounter{table}{-1}
\begin{table*}
\begin{center}
\caption{Continued.}
\begin{tabular}{l@{\hspace{0.1em}} c@{\hspace{.5em}} c@{\hspace{0.8em}} c@{\hspace{0.8em}} c@{\hspace{0.8em}} c@{\hspace{0.8em}} c@{\hspace{0.8em}} c@{\hspace{0.8em}} c@{\hspace{0.8em}} c@{\hspace{0.8em}} c@{\hspace{0.8em}} c@{\hspace{0.8em}} c@{\hspace{0.8em}} }
\hline \\
Cluster   & $ z $ & $ t_{\rm exp} $ & ID &   $T_{\rm ew}$ & $w$ & CC/NCC\\
          &        &        ks         & &  (keV)        & ($\times 10^{-2}$)  &\\
\hline \\
    SDSSJ0922+0345&   0.284 &     9.9&  11768& $   9.55\pm   2.32$ & $   2.83\pm   1.03$ &    NCC \\
    RXJ0232.2-4420&   0.284 &    22.6&   4993& $   9.02\pm   1.18$ & $   2.14\pm   0.78$ &     CC \\
             A1703&   0.284 &    29.1&  15123& $   5.95\pm   0.33$ & $   0.43\pm   0.16$ &    NCC \\
    RXJ0437.1+0043&   0.285 &    42.5&  11729  7900& $   8.10\pm   0.85$ & $   0.61\pm   0.22$ &     CC \\
          ZWCL3146&   0.290 &    82.0&    909  9371& $   9.98\pm   0.49$ & $   0.46\pm   0.17$ &     CC \\
             Z7215&   0.292 &    13.0&   7899& $   6.47\pm   1.18$ & $   2.08\pm   0.76$ &    NCC \\
             A2813&   0.292 &    19.9&   9409& $   6.45\pm   0.67$ & $   0.74\pm   0.27$ &    NCC \\
             A2537&   0.295 &    38.3&   9372& $   6.81\pm   0.72$ & $   0.81\pm   0.29$ &     CC \\
          ACOS0520&   0.295 &    31.0&   9331 15099& $   8.79\pm   1.41$ & $   1.86\pm   0.68$ &    NCC \\
            1E0657&   0.296 &   562.4&   3184  5356  5361   554  4984  4985  4986  5355& $  12.73\pm   0.26$ & $   3.10\pm   1.13$ &    NCC \\
 &  & &    5357  5358&  &\\
  ACT-CLJ0707-5522&   0.296 &    19.8&  12271& $   4.77\pm   0.44$ & $   2.52\pm   0.92$ &    NCC \\
              A781&   0.298 &    44.8&    534 15128& $   7.24\pm   0.71$ & $   3.67\pm   1.34$ &    NCC \\
     G292.51+21.98&   0.300 &    42.7&  15134& $   7.98\pm   0.81$ & $   0.83\pm   0.30$ &    NCC \\
            AS0295&   0.301 &    19.8&  12260& $   7.72\pm   0.86$ & $   2.08\pm   0.76$ &    NCC \\
     MS1008.1-1224&   0.301 &    51.2&    926  7711& $   7.49\pm   0.53$ & $   3.74\pm   1.37$ &    NCC \\
             A2552&   0.302 &    36.3&  11730  3288& $   9.30\pm   1.07$ & $   0.53\pm   0.19$ &    NCC \\
  MACSJ2245.0+2637&   0.304 &    14.2&   3287& $   6.93\pm   1.29$ & $   0.64\pm   0.23$ &     CC \\
             Z5699&   0.306 &    29.3&  14015& $   6.61\pm   0.84$ & $   1.53\pm   0.56$ &    NCC \\
  MACSJ1131.8-1955&   0.307 &    23.9&   3276 15300& $  10.51\pm   1.13$ & $   3.60\pm   1.31$ &    NCC \\
             A2744&   0.308 &   100.4&   7712  7915  8477  8557& $   9.33\pm   0.54$ & $   1.82\pm   0.66$ &    NCC \\
  MACSJ0242.5-2132&   0.314 &    11.6&   3266& $   5.87\pm   1.11$ & $   0.38\pm   0.14$ &     CC \\
   RXCJ2003.5-2323&   0.317 &    49.2&   7916& $   8.05\pm   0.59$ & $   5.61\pm   2.05$ &    NCC \\
  MACSJ1427.6-2521&   0.318 &    28.4&   9373& $   6.84\pm   0.89$ & $   0.36\pm   0.13$ &     CC \\
             A1995&   0.319 &   153.3&   7021  7713  7022  7023& $   6.79\pm   0.55$ & $   0.92\pm   0.34$ &    NCC \\
  SPT-CLJ2355-5055&   0.320 &    11.5&  11746& $   4.90\pm   0.65$ & $   1.05\pm   0.38$ &     CC \\
             A1351&   0.322 &    32.7&  15136& $   9.64\pm   1.09$ & $   3.49\pm   1.28$ &    NCC \\
  MACSJ0257.6-2209&   0.322 &    20.5&   3267& $   6.86\pm   1.13$ & $   0.65\pm   0.24$ &     CC \\
  MACSJ0308.9+2645&   0.324 &    22.7&   3268& $   9.59\pm   1.11$ & $   0.86\pm   0.31$ &     CC \\
  MACSJ2229.7-2755&   0.324 &    30.2&   3286  9374& $   5.14\pm   0.57$ & $   0.18\pm   0.06$ &     CC \\
  MACSJ2135.2-0102&   0.325 &    26.7&  11710& $   9.31\pm   1.77$ & $   1.39\pm   0.51$ &    NCC \\
  MACSJ2049.9-3217&   0.325 &    23.5&   3283& $   7.62\pm   1.07$ & $   1.70\pm   0.62$ &    NCC \\
     ZWCL1358+6245&   0.328 &     6.7&   7714& $   6.21\pm   1.24$ & $   0.55\pm   0.20$ &     CC \\
  MACSJ0712.3+5931&   0.328 &    25.7&  11709& $   6.39\pm   1.02$ & $   1.32\pm   0.48$ &     CC \\
  SPT-CLJ0236-4938&   0.334 &    39.6&  12266& $   5.40\pm   0.80$ & $   1.76\pm   0.64$ &    NCC \\
  MACSJ0520.7-1328&   0.340 &    18.5&   3272& $   8.07\pm   1.23$ & $   0.51\pm   0.19$ &     CC \\
  SPT-CLJ2031-4037&   0.342 &     9.9&  13517& $   6.51\pm   0.85$ & $   1.24\pm   0.45$ &     CC \\
  ACT-CLJ0217-5245&   0.343 &    19.8&  12269& $   6.37\pm   0.91$ & $   3.86\pm   1.41$ &    NCC \\
    RXJ1532.9+3021&   0.345 &    10.0&   1665& $   6.47\pm   0.99$ & $   0.14\pm   0.05$ &     CC \\
  SPT-CLJ0040-4407&   0.350 &     8.0&  13395& $   7.61\pm   0.93$ & $   0.87\pm   0.32$ &    NCC \\
  MACSJ1931.8-2634&   0.352 &   112.5&   3282  9382& $   7.92\pm   0.45$ & $   0.17\pm   0.06$ &     CC \\
  MACSJ1115.8+0129&   0.352 &    52.9&   3275  9375& $   9.14\pm   1.06$ & $   0.36\pm   0.13$ &     CC \\
            A1063S&   0.354 &    26.7&   4966& $  11.54\pm   1.11$ & $   1.43\pm   0.52$ &     CC \\
           RBS0797&   0.354 &    11.7&   2202& $   9.94\pm   1.29$ & $   0.21\pm   0.08$ &     CC \\
  MACSJ0404.6+1109&   0.355 &    21.3&   3269& $   6.59\pm   1.16$ & $   2.07\pm   0.75$ &    NCC \\
  SPT-CLJ2325-4111&   0.358 &     8.9&  13405& $   6.50\pm   1.17$ & $   3.23\pm   1.18$ &    NCC \\
  SPT-CLJ0348-4515&   0.358 &    12.6&  13465& $   5.48\pm   0.96$ & $   1.57\pm   0.57$ &    NCC \\
  MACSJ0011.7-1523&   0.360 &    58.9&   3261  6105& $   6.21\pm   0.49$ & $   0.53\pm   0.19$ &     CC \\
  MACSJ0150.3-1005&   0.363 &    26.8&  11711& $   6.75\pm   0.85$ & $   0.39\pm   0.14$ &     CC \\
  MACSJ0035.4-2015&   0.364 &    21.4&   3262& $   8.43\pm   0.56$ & $   0.51\pm   0.18$ &     CC \\
    RXJ0027.6+2616&   0.367 &    31.4&   3249 14012& $   7.24\pm   1.34$ & $   1.05\pm   0.38$ &    NCC \\
      CLJ0318-0302&   0.370 &    14.3&   5775& $   4.25\pm   0.45$ & $   3.44\pm   1.26$ &    NCC \\
  SPT-CLJ0411-4819&   0.370 &     8.0&  13396& $   5.63\pm   1.01$ & $   0.99\pm   0.36$ &     CC \\
          ZWCL1953&   0.380 &    28.1&   1659  7716& $   8.35\pm   1.11$ & $   1.54\pm   0.56$ &    NCC \\
  MACSJ0949.8+1708&   0.384 &    14.3&   3274& $   9.52\pm   1.87$ & $   1.50\pm   0.55$ &    NCC \\
  MACSJ1731.6+2252&   0.389 &    20.5&   3281& $   8.97\pm   1.74$ & $   1.87\pm   0.68$ &    NCC \\
    RXJ1720.2+3536&   0.391 &    63.2&   3280  6107  7225  7718& $   6.50\pm   0.43$ & $   0.60\pm   0.22$ &     CC \\
  SPT-CLJ0304-4921&   0.392 &    20.8&  12265& $   6.93\pm   1.25$ & $   0.68\pm   0.25$ &     CC \\
  MACSJ1354.6+7715&   0.396 &    32.6&  11754& $   6.86\pm   1.11$ & $   1.15\pm   0.42$ &     CC \\
  MACSJ0416.1-2403&   0.397 &    15.8&  10446& $   8.14\pm   1.23$ & $   0.59\pm   0.22$ &    NCC \\
  MACSJ0429.6-0253&   0.399 &    22.9&   3271& $   8.55\pm   1.49$ & $   0.47\pm   0.17$ &     CC \\
\end{tabular}
\end{center}
\end{table*}

\addtocounter{table}{-1}
\begin{table*}
\begin{center}
\caption{Continued.}
\begin{tabular}{l@{\hspace{0.1em}} c@{\hspace{.5em}} c@{\hspace{0.8em}} c@{\hspace{0.8em}} c@{\hspace{0.8em}} c@{\hspace{0.8em}} c@{\hspace{0.8em}} c@{\hspace{0.8em}} c@{\hspace{0.8em}} c@{\hspace{0.8em}} c@{\hspace{0.8em}} c@{\hspace{0.8em}} c@{\hspace{0.8em}} }
\hline \\
Cluster   & $ z $ & $ t_{\rm exp} $ & ID &   $T_{\rm ew}$ & $w$ & CC/NCC\\
          &        &        ks         & &  (keV)        & ($\times 10^{-2}$)  &\\
\hline \\
  MACSJ0159.8-0849&   0.406 &    72.5&   3265  6106  9376& $   8.85\pm   0.87$ & $   0.66\pm   0.24$ &     CC \\
  SPT-CLJ0013-4906&   0.406 &    14.1&  13462& $   6.12\pm   1.09$ & $   1.14\pm   0.42$ &    NCC \\
      CLJ1213+0253&   0.409 &    18.4&   4934& $   4.29\pm   0.88$ & $   1.53\pm   0.56$ &    NCC \\
    MACSJ2228+2036&   0.412 &    19.6&   3285& $   9.20\pm   1.45$ & $   2.63\pm   0.96$ &    NCC \\
  SPT-CLJ0234-5831&   0.415 &    10.0&  13403& $   8.11\pm   1.05$ & $   0.49\pm   0.18$ &     CC \\
    SLJ1602.8+4335&   0.415 &    42.3&  12308& $   5.58\pm   1.61$ & $   0.66\pm   0.24$ &    NCC \\
  SPT-CLJ0555-6405&   0.420 &    10.9&  13404& $   8.65\pm   2.04$ & $   2.67\pm   0.97$ &    NCC \\
  MACSJ2046.0-3430&   0.420 &    39.2&   9377& $   5.73\pm   0.71$ & $   0.24\pm   0.09$ &     CC \\
  SPT-CLJ0252-4824&   0.421 &    30.8&  13494& $   6.74\pm   0.90$ & $   5.59\pm   2.04$ &    NCC \\
  SPT-CLJ0438-5419&   0.422 &    19.8&  12259& $  13.17\pm   3.18$ & $   1.51\pm   0.55$ &     CC \\
  SPT-CLJ0551-5709&   0.423 &    36.1&  11743 11871& $   5.03\pm   0.88$ & $   2.47\pm   0.90$ &    NCC \\
     MS0302.7+1658&   0.424 &    10.0&    525& $   3.94\pm   0.50$ & $   2.07\pm   0.75$ &     CC \\
     MS1621.5+2640&   0.426 &    29.6&    546& $   7.18\pm   1.38$ & $   1.70\pm   0.62$ &    NCC \\
  SPT-CLJ2135-5726&   0.427 &    17.1&  13463& $   6.93\pm   1.18$ & $   0.77\pm   0.28$ &    NCC \\
      CLJ1216+2633&   0.428 &    18.5&   4931& $   5.90\pm   1.80$ & $   1.15\pm   0.42$ &    NCC \\
  MACSJ0358.8-2955&   0.428 &    57.8&  11719 12300 13194& $   8.00\pm   0.72$ & $   1.88\pm   0.69$ &     CC \\
  MACSJ0451.9+0006&   0.430 &    10.0&   5815& $   7.66\pm   1.04$ & $   2.01\pm   0.73$ &    NCC \\
  MACSJ0553.4-3342&   0.431 &    82.9&  12244  5813& $  12.68\pm   1.26$ & $   2.76\pm   1.01$ &    NCC \\
  MACSJ1226.8+2153&   0.436 &   129.8&  12878& $   6.09\pm   0.52$ & $   0.38\pm   0.14$ &    NCC \\
  MACSJ1115.2+5320&   0.439 &    33.6&   3253  5008  5350& $   8.59\pm   1.10$ & $   2.24\pm   0.82$ &    NCC \\
  MACSJ0417.5-1154&   0.440 &    89.1&   3270 11759 12010& $  10.65\pm   0.54$ & $   2.77\pm   1.01$ &     CC \\
  MACSJ1206.2-0847&   0.440 &    23.2&   3277& $  11.41\pm   1.64$ & $   1.31\pm   0.48$ &     CC \\
  MACSJ0913.7+4056&   0.442 &    76.1&  10445& $   6.86\pm   0.88$ & $   0.22\pm   0.08$ &     CC \\
  SPT-CLJ0330-5228&   0.442 &    19.4&    893& $   4.75\pm   0.49$ & $   1.92\pm   0.70$ &    NCC \\
  MACSJ2243.3-0935&   0.447 &    20.0&   3260& $   9.74\pm   1.26$ & $   0.96\pm   0.35$ &    NCC \\
  MACSJ0329.6-0211&   0.450 &    69.4&   3257  3582  6108& $   5.85\pm   0.45$ & $   0.83\pm   0.30$ &     CC \\
  MACSJ1359.1-1929&   0.450 &    49.2&   9378& $   7.38\pm   1.35$ & $   0.76\pm   0.28$ &     CC \\
    RXJ1347.5-1145&   0.451 &   213.8&   3592 13516 13999 14407& $  13.63\pm   0.69$ & $   0.84\pm   0.31$ &     CC \\
  MACSJ0140.0-0555&   0.451 &    29.5&   5013 12243& $   8.60\pm   1.56$ & $   2.41\pm   0.88$ &    NCC \\
      RXJ1701+6414&   0.453 &    49.3&    547& $   6.39\pm   0.72$ & $   1.59\pm   0.58$ &     CC \\
  SPT-CLJ0509-5342&   0.461 &    28.8&   9432& $   6.49\pm   1.21$ & $   0.59\pm   0.22$ &     CC \\
  MACSJ1621.3+3810&   0.461 &   155.0&   3254  3594  6109  6172  9379 10785& $   7.41\pm   0.67$ & $   0.37\pm   0.14$ &     CC \\
      CLJ1641+4001&   0.464 &    44.5&   3575& $   5.03\pm   0.68$ & $   4.17\pm   1.52$ &    NCC \\
             3C295&   0.464 &    87.7&   2254& $   5.65\pm   0.68$ & $   1.64\pm   0.60$ &     CC \\
  SPT-CLJ2035-5251&   0.470 &    18.4&  13466& $   4.90\pm   0.84$ & $   0.59\pm   0.22$ &    NCC \\
  SPT-CLJ0655-5234&   0.470 &    20.2&  13486& $   8.58\pm   1.91$ & $   1.33\pm   0.49$ &    NCC \\
      CLJ0522-3625&   0.472 &    45.3&   4926  5837& $   4.94\pm   0.66$ & $   1.42\pm   0.52$ &    NCC \\
      CLJ0853+5759&   0.475 &    42.4&   5765  4925& $   7.51\pm   1.85$ & $   1.45\pm   0.53$ &    NCC \\
  SPT-CLJ2145-5644&   0.480 &    14.9&  13398& $   7.59\pm   1.07$ & $   0.69\pm   0.25$ &    NCC \\
  ACT-CLJ0215-5212&   0.480 &    20.8&  12268& $   5.02\pm   1.05$ & $   1.52\pm   0.55$ &    NCC \\
  SPT-CLJ2233-5339&   0.480 &    17.1&  13504& $   8.50\pm   1.27$ & $   0.89\pm   0.32$ &    NCC \\
  SPT-CLJ0334-4659&   0.485 &    25.7&  13470& $   6.44\pm   0.63$ & $   1.13\pm   0.41$ &     CC \\
  MACSJ1824.3+4309&   0.487 &    14.9&   3255& $   4.74\pm   0.97$ & $   3.99\pm   1.46$ &     CC \\
      CLJ0926+1242&   0.489 &    50.0&   4929  5838& $   4.79\pm   0.56$ & $   0.56\pm   0.20$ &    NCC \\
  MACSJ1427.3+4408&   0.490 &    40.8&   9380  9808& $   8.94\pm   1.05$ & $   0.38\pm   0.14$ &     CC \\
  MACSJ1311.0-0310&   0.494 &   114.6&   7721  3258  6110  9381& $   6.36\pm   0.58$ & $   0.48\pm   0.18$ &     CC \\
  SPT-CLJ0200-4852&   0.498 &    23.5&  13487& $   7.77\pm   1.36$ & $   1.82\pm   0.67$ &    NCC \\
      CLJ0030+2618&   0.500 &    16.1&   5762& $   5.03\pm   0.97$ & $   1.22\pm   0.44$ &    NCC \\
  MACSJ2214.9-1359&   0.503 &    32.9&   3259  5011& $  10.04\pm   1.37$ & $   0.91\pm   0.33$ &    NCC \\
  MACSJ0257.1-2325&   0.505 &    38.1&   1654  3581& $  10.41\pm   1.62$ & $   0.70\pm   0.26$ &     CC \\
      RXJ1525+0958&   0.516 &    49.1&   1664& $   5.18\pm   0.44$ & $   1.52\pm   0.55$ &    NCC \\
  SPT-CLJ0304-4401&   0.520 &    14.9&  13402& $   7.43\pm   1.42$ & $   0.95\pm   0.35$ &    NCC \\
  SPT-CLJ0346-5439&   0.530 &    17.8&  12270& $   5.15\pm   0.69$ & $   1.23\pm   0.45$ &    NCC \\
  ACT-CLJ0346-5438&   0.530 &    16.1&  13155& $   4.85\pm   0.67$ & $   0.99\pm   0.36$ &    NCC \\
        X0916+2950&   0.530 &    38.3&  12913& $   2.81\pm   0.76$ & $   1.82\pm   0.67$ &    NCC \\
  SPT-CLJ2306-6505&   0.530 &    25.3&  13503& $   5.62\pm   0.95$ & $   0.93\pm   0.34$ &    NCC \\
     MS0016.9+1609&   0.541 &    67.4&    520& $   9.26\pm   0.77$ & $   0.46\pm   0.17$ &    NCC \\
  MACSJ1149.5+2223&   0.544 &    38.3&   1656  3589& $   8.76\pm   1.05$ & $   1.48\pm   0.54$ &    NCC \\
    RXJ1423.8+2404&   0.545 &    18.5&   1657& $   6.81\pm   0.81$ & $   0.37\pm   0.13$ &     CC \\
      CLJ1354-0221&   0.546 &    54.3&   5835  4932& $   5.78\pm   0.98$ & $   1.64\pm   0.60$ &    NCC \\
   
\end{tabular}
\end{center}
\end{table*}

\addtocounter{table}{-1}
\begin{table*}
\begin{center}
\caption{Continued.}
\begin{tabular}{l@{\hspace{0.1em}} c@{\hspace{.5em}} c@{\hspace{0.8em}} c@{\hspace{0.8em}} c@{\hspace{0.8em}} c@{\hspace{0.8em}} c@{\hspace{0.8em}} c@{\hspace{0.8em}} c@{\hspace{0.8em}} c@{\hspace{0.8em}} c@{\hspace{0.8em}} c@{\hspace{0.8em}} c@{\hspace{0.8em}} }
\hline \\
Cluster   & $ z $ & $ t_{\rm exp} $ & ID &   $T_{\rm ew}$ & $w$ & CC/NCC\\
          &        &        ks         & &  (keV)        & ($\times 10^{-2}$)  &\\
\hline \\
  SPT-CLJ2335-4544&   0.547 &    20.7&  13496& $   7.67\pm   1.29$ & $   1.27\pm   0.46$ &    NCC \\
  MACSJ0717.5+3745&   0.548 &   149.1&   1655  4200 16235& $  10.55\pm   0.55$ & $   1.60\pm   0.59$ &    NCC \\
      CLJ1117+1744&   0.548 &    63.4&   4933  5836& $   4.47\pm   0.88$ & $   1.59\pm   0.58$ &    NCC \\
  SPT-CLJ0307-5042&   0.550 &    39.4&  13476& $   7.65\pm   1.19$ & $   0.48\pm   0.17$ &    NCC \\
  SPT-CLJ0232-5257&   0.556 &    19.8&  12263& $   6.87\pm   1.19$ & $   3.54\pm   1.29$ &    NCC \\
      RXJ1121+2327&   0.562 &    69.2&   1660& $   4.24\pm   0.33$ & $   1.30\pm   0.48$ &    NCC \\
  SPT-CLJ0456-5116&   0.562 &    49.4&  13474& $   5.98\pm   0.95$ & $   1.15\pm   0.42$ &    NCC \\
  SPT-CLJ2148-6116&   0.571 &    36.0&  13488& $   7.34\pm   1.19$ & $   1.75\pm   0.64$ &    NCC \\
    CLJ0848.7+4456&   0.574 &   184.8&   1708   927& $   2.72\pm   0.35$ & $   1.67\pm   0.61$ &    NCC \\
  SPT-CLJ2331-5051&   0.576 &    28.7&   9333& $   4.80\pm   0.86$ & $   0.54\pm   0.20$ &     CC \\
      CLJ0216-1747&   0.578 &    63.1&   6393  5760& $   4.24\pm   0.71$ & $   1.52\pm   0.55$ &    NCC \\
  SPT-CLJ0307-6225&   0.579 &    24.7&  12191& $   5.11\pm   0.79$ & $   2.21\pm   0.81$ &    NCC \\

  SPT-CLJ2245-6206&   0.580 &    29.3&  13499& $   5.95\pm   1.02$ & $   5.72\pm   2.56$ &    NCC \\
     MS2053.7-0449&   0.583 &    88.6&   1667   551& $   4.57\pm   0.68$ & $   1.01\pm   0.37$ &    NCC \\
    RXJ0647.7+7015&   0.584 &    39.3&   3196  3584& $  12.57\pm   2.10$ & $   1.10\pm   0.40$ &    NCC \\
  MACSJ0025.4-1222&   0.585 &   156.9&  10413 10786 10797  3251  5010& $   7.94\pm   0.54$ & $   1.34\pm   0.49$ &    NCC \\
      CLJ0956+4107&   0.587 &    58.4&   5759  4930  5294& $   5.23\pm   0.85$ & $   2.13\pm   0.78$ &    NCC \\
  MACSJ2129.4-0741&   0.594 &    37.8&   3595  3199& $   8.11\pm   0.89$ & $   2.84\pm   1.04$ &    NCC \\
  SPT-CLJ2232-5959&   0.594 &    31.9&  13502& $   5.20\pm   0.55$ & $   0.65\pm   0.24$ &     CC \\
  SPT-CLJ2344-4243&   0.595 &    11.9&  13401& $  13.68\pm   4.24$ & $   0.21\pm   0.08$ &     CC \\
  SPT-CLJ0033-6326&   0.597 &    21.1&  13483& $   7.60\pm   2.30$ & $   1.65\pm   0.60$ &    NCC \\
      CLJ1120+4318&   0.600 &    19.8&   5771& $   4.80\pm   0.52$ & $   1.14\pm   0.42$ &    NCC \\
  SPT-CLJ0559-5249&   0.609 &   108.4&  12264 13116 13117& $   5.77\pm   0.76$ & $   1.84\pm   0.67$ &    NCC \\
      CLJ1334+5031&   0.620 &    19.5&   5772& $   4.99\pm   1.35$ & $   0.77\pm   0.28$ &    NCC \\
  SPT-CLJ0417-4748&   0.620 &    21.8&  13397& $   6.55\pm   0.78$ & $   0.64\pm   0.23$ &     CC \\
  SPT-CLJ0123-4821&   0.620 &    71.1&  13491& $   5.78\pm   0.89$ & $   1.52\pm   0.56$ &    NCC \\
  SPT-CLJ0256-5617&   0.630 &    47.4&  13481 14448& $   7.77\pm   0.99$ & $   2.44\pm   0.89$ &    NCC \\
  SPT-CLJ0426-5455&   0.630 &    32.4&  13472& $   5.50\pm   1.04$ & $   2.02\pm   0.74$ &    NCC \\
    CLJ0542.8-4100&   0.634 &    50.4&    914& $   7.37\pm   1.37$ & $   2.40\pm   0.88$ &    NCC \\
  SPT-CLJ0243-5930&   0.650 &    46.9&  13484 15573& $   7.69\pm   0.88$ & $   1.41\pm   0.51$ &    NCC \\
  SPT-CLJ2218-4519&   0.650 &    34.8&  13501& $   8.45\pm   1.53$ & $   2.07\pm   0.76$ &    NCC \\
  SPT-CLJ2222-4834&   0.652 &    32.5&  13497& $   6.02\pm   0.87$ & $   1.16\pm   0.43$ &     CC \\
  SPT-CLJ0212-4657&   0.655 &    28.0&  13464& $   3.84\pm   0.61$ & $   3.50\pm   1.43$ &    NCC \\
  SPT-CLJ0352-5647&   0.660 &    44.4&  13490 15571& $   5.76\pm   0.87$ & $   0.94\pm   0.34$ &    NCC \\
  SPT-CLJ0616-5227&   0.684 &    38.6&  12261 13127& $   6.81\pm   0.95$ & $   1.96\pm   0.72$ &    NCC \\
  MACSJ0744.8+3927&   0.686 &    87.3&   3197  3585  6111& $   8.90\pm   0.97$ & $   1.36\pm   0.50$ &     CC \\
           RXJ1757&   0.690 &    46.1&  10443 11999& $   4.12\pm   1.14$ & $   1.15\pm   0.42$ &    NCC \\
       Cl1324+3059&   0.696 &    48.4&   9403  9840& $   4.13\pm   2.16$ & $   2.21\pm   0.81$ &    NCC \\
    RCS2327.4-0204&   0.700 &   141.0&  14025 14361& $  10.60\pm   0.92$ & $   0.29\pm   0.11$ &     CC \\
      RXJ1221+4918&   0.700 &    78.3&   1662& $   8.10\pm   1.25$ & $   2.18\pm   0.79$ &    NCC \\
  SPT-CLJ0000-5748&   0.702 &    30.1&   9335& $   6.54\pm   0.76$ & $   0.92\pm   0.34$ &     CC \\
  SPT-CLJ0310-4647&   0.709 &    36.7&  13492& $   7.21\pm   1.38$ & $   1.05\pm   0.38$ &    NCC \\
  SPT-CLJ0102-4603&   0.720 &    60.7&  13485& $   4.56\pm   0.80$ & $   4.56\pm   1.66$ &    NCC \\
    CLJ2302.8+0844&   0.722 &   104.7&    918& $   6.89\pm   1.11$ & $   5.63\pm   2.06$ &    NCC \\
  SPT-CLJ2043-5035&   0.723 &    79.4&  13478& $   6.72\pm   0.74$ & $   0.45\pm   0.17$ &     CC \\
    CLJ1113.1-2615&   0.725 &   102.3&    915& $   5.22\pm   0.66$ & $   2.02\pm   0.74$ &    NCC \\
  SPT-CLJ0142-5032&   0.730 &    29.1&  13467& $   8.92\pm   2.22$ & $   0.89\pm   0.32$ &    NCC \\
  SPT-CLJ2301-4023&   0.730 &    57.7&  13505& $   5.95\pm   0.90$ & $   0.89\pm   0.32$ &    NCC \\
  SPT-CLJ2352-4657&   0.730 &    79.4&  13506& $   8.03\pm   1.73$ & $   3.77\pm   1.38$ &    NCC \\
  SPT-CLJ0406-4805&   0.737 &    26.0&  13477& $   6.50\pm   1.27$ & $   3.83\pm   1.40$ &    NCC \\
  SPT-CLJ0324-6236&   0.740 &    54.2&  12181 13137 13213& $   7.29\pm   1.20$ & $   1.42\pm   0.52$ &    NCC \\
  SPT-CLJ2259-6057&   0.750 &    64.0&  13498& $   6.32\pm   0.87$ & $   1.96\pm   0.72$ &    NCC \\
  SPT-CLJ0014-4952&   0.752 &    55.0&  13471& $   7.17\pm   0.85$ & $   6.79\pm   2.48$ &    NCC \\
       Cl1324+3011&   0.755 &    50.2&   9404  9836& $   4.69\pm   2.30$ & $   1.61\pm   0.59$ &    NCC \\
  SPT-CLJ0528-5300&   0.768 &   124.0&  11747 11874 12092 13126  9341 10862 11996& $   5.26\pm   0.78$ & $   0.78\pm   0.28$ &    NCC \\
  SPT-CLJ2337-5942&   0.775 &    19.7&  11859& $   9.12\pm   1.52$ & $   0.77\pm   0.28$ &    NCC \\
  SPT-CLJ2359-5009&   0.775 &   131.1&   9334 11742 11864 11997& $   6.45\pm   1.08$ & $   1.06\pm   0.39$ &    NCC \\
      RCS2318+0034&   0.780 &   149.5&  11718& $   7.89\pm   1.74$ & $   1.06\pm   0.39$ &    NCC \\
  SPT-CLJ0449-4901&   0.790 &    51.0&  13473& $   8.21\pm   1.43$ & $   2.91\pm   1.06$ &    NCC \\
  SPT-CLJ0441-4855&   0.790 &    69.2&  13475 14371 14372& $   7.09\pm   0.92$ & $   0.92\pm   0.34$ &    NCC \\
  
\end{tabular}
\end{center}
\end{table*}

\addtocounter{table}{-1}
\begin{table*}
\begin{center}
\caption{Continued.}
\begin{tabular}{l@{\hspace{0.1em}} c@{\hspace{.5em}} c@{\hspace{0.8em}} c@{\hspace{0.8em}} c@{\hspace{0.8em}} c@{\hspace{0.8em}} c@{\hspace{0.8em}} c@{\hspace{0.8em}} c@{\hspace{0.8em}} c@{\hspace{0.8em}} c@{\hspace{0.8em}} c@{\hspace{0.8em}} c@{\hspace{0.8em}} }
\hline \\
Cluster   & $ z $ & $ t_{\rm exp} $ & ID &   $T_{\rm ew}$ & $w$ & CC/NCC\\
          &        &        ks         & &  (keV)        & ($\times 10^{-2}$)  &\\
\hline \\
   RXJ1350.0+6007&   0.804 &    57.8&   2229& $   4.79\pm   0.79$ & $   2.63\pm   0.96$ &    NCC \\
    RXJ1317.4+2911&   0.805 &   110.3&   2228& $   4.72\pm   1.56$ & $   3.49\pm   1.27$ &    NCC \\
    RXJ1716.9+6708&   0.813 &    51.2&    548& $   6.93\pm   1.08$ & $   1.18\pm   0.43$ &    NCC \\
    RXJ1821.6+6827&   0.820 &    49.6&  10444 10924& $   4.70\pm   0.81$ & $   1.85\pm   0.68$ &    NCC \\
  SPT-CLJ0058-6145&   0.830 &    50.6&  13479& $   5.76\pm   1.40$ & $   1.06\pm   0.39$ &    NCC \\
  SPT-CLJ2258-4044&   0.830 &    54.4&  13495& $   7.30\pm   1.07$ & $   1.80\pm   0.66$ &    NCC \\
    CLJ0152.7-1357&   0.831 &    35.4&    913& $   4.85\pm   0.95$ & $   2.70\pm   0.99$ &    NCC \\
  SPT-CLJ0102-4915&   0.870 &    59.2&  12258& $  13.17\pm   1.35$ & $   7.10\pm   2.59$ &     CC \\
  SPT-CLJ0533-5005&   0.881 &    71.7&  11748 12001 12002& $   6.11\pm   2.05$ & $   1.72\pm   0.63$ &    NCC \\
    CLJ1226.9+3332&   0.890 &    63.4&   3180  5014& $  11.68\pm   1.73$ & $   1.09\pm   0.40$ &     CC \\
            Cl1604&   0.898 &    94.5&   6932  6933  7343& $   4.30\pm   2.07$ & $   1.83\pm   0.67$ &    NCC \\
  SPT-CLJ2034-5936&   0.920 &    58.7&  12182& $   7.92\pm   1.35$ & $   1.13\pm   0.41$ &    NCC \\
  SPT-CLJ2146-4632&   0.932 &    81.0&  13469& $   5.24\pm   0.62$ & $   1.58\pm   0.58$ &    NCC \\
   PLCKG266.6-27.3&   0.940 &   240.5&  14017 14018 14349 14350 14351 14437 15572 15574& $  10.32\pm   0.50$ & $   0.77\pm   0.28$ &     CC \\
 &  & &   15579 15582 15588 15589&  &\\

  SPT-CLJ2345-6405&   0.940 &    65.2&  13500& $   6.59\pm   0.92$ & $   1.66\pm   0.61$ &    NCC \\
  SPT-CLJ2341-5119&   1.000 &    80.0&  11799  9345& $   6.32\pm   0.85$ & $   1.47\pm   0.54$ &    NCC \\
  SPT-CLJ0037-5047&   1.026 &    72.5&  13493& $   3.22\pm   1.05$ & $   3.86\pm   1.41$ &    NCC \\
    CLJ1415.1+3612&   1.030 &    88.6&   4163& $   6.74\pm   0.69$ & $   1.05\pm   0.38$ &    NCC \\
  SPT-CLJ0151-5954&   1.030 &    49.4&  13480& $   4.02\pm   0.62$ & $   1.53\pm   0.56$ &    NCC \\
  SPT-CLJ0546-5345&   1.066 &    69.5&   9332  9336 10851 10864 11739& $   8.54\pm   1.11$ & $   1.02\pm   0.37$ &    NCC \\
  SPT-CLJ2342-5411&   1.075 &   174.0&  11741 11870 12014 12091& $   4.70\pm   0.61$ & $   1.43\pm   0.52$ &     CC \\
      RXJ0910+5422&   1.110 &   169.8&   2452  2227& $   4.58\pm   1.04$ & $   1.54\pm   0.56$ &    NCC \\
  SPT-CLJ2106-5844&   1.132 &    24.7&  12180& $   8.80\pm   1.44$ & $   1.45\pm   0.53$ &    NCC \\
  SPT-CLJ0446-5849&   1.160 &    53.4&  13482 15560& $   6.13\pm   2.03$ & $   2.44\pm   0.89$ &    NCC \\
  SPT-CLJ2236-4555&   1.160 &    83.0&  13507 15266& $   7.44\pm   1.39$ & $   2.85\pm   1.04$ &    NCC \\
  SPT-CLJ0156-5541&   1.220 &    77.7&  13489& $   7.81\pm   1.23$ & $   0.75\pm   0.28$ &    NCC \\
      RDCSJ1252-29&   1.237 &   187.3&   4198  4403& $   6.50\pm   1.08$ & $   1.52\pm   0.56$ &    NCC \\
 
\hline \\ 
\end{tabular}
\end{center}
\end{table*}

\end{document}